\documentclass[aps,prd,nofootinbib,superscriptaddress,onecolumn,notitlepage,longbibliography]{revtex4-1}

\usepackage{amsmath,graphicx,verbatim,epsfig}
\usepackage{epstopdf}
\usepackage[utf8x]{inputenc}
\usepackage{amsmath,graphicx}
\usepackage[normalem]{ulem}
\usepackage[colorlinks=true,linkcolor=blue,citecolor=blue]{hyperref}

\newcommand{\nn}{\nonumber}

\newcommand{\bn}{{\bar n}}

\newcommand{\bnslash}{{\not \!\bn}}

\newcommand{\be}{\begin{equation}}
\newcommand{\ee}{\end{equation}}
\newcommand{\bea}{\begin{eqnarray}}
\newcommand{\eea}{\end{eqnarray}}
\newcommand{\balign}{\begin{align}}
\newcommand{\ealign}{\end{align}}
\newcommand{\as}{\alpha_s}

\newcommand{\sandwich}[3]{\left< #1 \right | #2 \left | #3 \right >}

\newcommand{\bg}{\begin{gather}}
\newcommand{\foma}{\end{gather}}

\newcommand{\noopsort}[1]{}

\def\L{\Lambda}

\def\z{\zeta}

\def\<{\langle}
\def\>{\rangle}

\def\a{\alpha}
\def\b{\beta}
\def\g{\gamma}  \def\G{\Gamma}
  \def\D{\Delta}
   \def\L{\Lambda}
\def\s{\sigma}
  
\def\x{\xi}

\def\m{\mu}

\def\z{\zeta}

\def\({\left(}
\def\[{\left[}
\def\){\right)}
\def\]{\right]}

\def\ln{\hbox{ln}}

\def\bnslash{\bar n\!\!\!\slash}

\def \le { \left    }
\def \ri { \right }

\def\lqcd{\L_{\rm QCD}}

\begin{document}


\title{Model Independent Evolution of Transverse Momentum Dependent Distribution Functions (TMDs) at NNLL}

\author{Miguel G. Echevarr\'ia}
\email{miguel.gechevarria@fis.ucm.es}
\affiliation{Departamento de F\'isica Te\'orica II,
Universidad Complutense de Madrid (UCM),
28040 Madrid, Spain}

\author{Ahmad Idilbi}
\email{idilbi@ectstar.eu}
\affiliation{European Centre for Theoretical Studies in Nuclear Physics and Related Areas (ECT*),
Villa Tambosi, Strada delle Tabarelle 286, I-38123, Villazzano, Trento, Italy}

\author{Andreas Sch\"afer}
\email{andreas.schaefer@physik.uni-regensburg.de}
\affiliation{Instit\"{u}t f\"{u}r Theoretische Physik, Universit\"at Regensburg,
D-93040 Regensburg, Germany}

\author{Ignazio Scimemi}
\email{ignazios@fis.ucm.es}
\affiliation{Departamento de F\'isica Te\'orica II,
Universidad Complutense de Madrid (UCM),
28040 Madrid, Spain}




\begin{abstract}

We discuss the evolution of the eight leading twist transverse momentum dependent parton distribution functions, which turns out to be universal and spin independent.
By using the highest order perturbatively calculable ingredients at our disposal, we perform the resummation of the large logarithms that appear in the evolution kernel of transverse momentum distributions up to next-to-next-to-leading logarithms (NNLL), thus obtaining an expression for the kernel with highly reduced model dependence. 
Our results can also be obtained using the standard CSS approach when a particular choice of the $b^*$ prescription is used. 
In this sense, and while restricted to the perturbative domain of applicability, we consider our results as a ``prediction'' of the correct value of $b_{\rm max}$ which is very close to $1.5~{\rm GeV}^{-1}$.
We explore under which kinematical conditions the effects of the non-perturbative region are negligible, and hence the evolution of transverse momentum distributions can be applied in a model independent way.
The application of the kernel is illustrated by considering the unpolarized transverse momentum dependent parton distribution function and the Sivers function.

\end{abstract}


\maketitle


\section{Introduction}

Transverse momentum distributions (TMDs) are needed for all processes for which intrinsic transverse parton momenta are relevant, which form a large group.
For example spin-dependent transverse momentum asymmetries provide unique clues to clarify the internal spin, angular momentum and 3-dimensional structure of hadrons.
While operating at different energies, experiments and facilities such as HERMES, COMPASS, JLab, Belle and BNL are pursuing intensive research programs to explore TMDs.
For the projected EIC, TMDs would be actually in the focus of the attention~\cite{Accardi:2012hwp}, and therefore there is a real need to sharpen the available theoretical tools for their extraction from data.
On the theoretical side, Sivers \cite{Sivers:1989cc} and Collins \cite{Collins:1992kk} asymmetries have been intensely studied (see \cite{Barone:2010zz} for a review of TMD functions in spin-physics), and have attracted much attention recently \cite{Aybat:2011zv,Aybat:2011ge,Aybat:2011ta,Kang:2011mr,Anselmino:2012aa,
Sun:2011iw,Musch:2011er}.
Basically, some of the observed spin-asymmetries are linked to the presence of gauge links in non-local correlators needed to maintain gauge invariance.

Based on the approach to TMDs developed by three of us in~\cite{Echevarria:2012js}, which is a generalization of the one given in~\cite{GarciaEchevarria:2011rb,Echevarria:2012qe}, and the one of Collins~\cite{Collins:2011zzd} (see also~\cite{Collins:2012uy}), in this paper we focus on the evolution kernel for TMDs.
Using the recently extracted anomalous dimension of the unpolarized quark-TMDPDF up to ${\cal O}(\alpha_s^3)$~\cite{Echevarria:2012js,GarciaEchevarria:2011rb}, and motivated by effective field theory methodology, below we offer a method to resum the large logarithms that appear in this kernel up to the highest possible logarithmic accuracy (NNLL).
As this kernel is the same for all eight leading twist TMDs, we discuss under which conditions it can be applied in a model independent way to extract them from data.

The study of the unpolarized TMDs was pioneered by Collins and Soper~\cite{Collins:1981uw,Collins:1981uk}.
Collins' new approach to TMDs~\cite{Collins:2011zzd} is based on defining those quantities in a way consistent with a generic factorization theorem, extracting their anomalous dimensions and their evolution properties.
This approach relies mainly on taking some of the Wilson lines in the soft factor off-the-light-cone.
When doing so, one introduces an auxiliary parameter $\zeta$, which specifies the measure of ``off-the-light-coness''.
A differential equation with respect to $\zeta$, the Collins-Soper evolution equation, is then derived and solved to resum large logarithms and determine the evolution of the non-perturbative TMDs with energy.
The resummation of the Collins-Soper kernel is done following the Collins-Soper-Sterman (CSS) method~\cite{Collins:1984kg}, which is based on using an effective strong coupling. 
This in turn leads to the emergence of a divergence when the coupling constant hits the Landau pole, an issue which is then avoided by the introduction of an smooth cutoff through the $b^*$ prescription and a non-perturbative model.
The value of $b_{\rm max}$ and the parameters of the model can only be extracted through fitting a resummed cross section to experimental data.

As will be discussed at length in section~\ref{sec:comparison} and appendices~\ref{app:CSS} and~\ref{sec:uyy}, our approach is not in contradiction with the standard CSS one. 
We present, however, a resummation scheme of which we hope that it reduces the remaining sensitivity to unknown non-perturbative contributions for practical fits to data in comparison to the CSS approach. 
The latter introduces a $b^*$ prescription which leads to a specific mixture of perturbative and non-perturbative contributions, of which those authors hope that it will on average give the best numerical control when fitting data. 
In contrast, we prefer to avoid such a mixture and try to clearly separate a resummed perturbative contribution from a remaining genuine non-perturbative one, specifying under which kinematical conditions the latter is negligible.
As usual for such resummation schemes, the practical value of our approach can only be judged by its phenomenological success.

Comparing with the standard CSS approach and the already existing fits of the non-perturbative Brock-Landry-Nadolsky-Yuan (BLNY) model, we find that the phenomenologically preferred value of the crucial cut-off parameter of that scheme $b_{\rm max}=1.5~{\rm GeV}^{-1}$, see~\cite{Konychev:2005iy}, is actually strongly suggested by our new scheme, which is parameter free in the perturbative domain. 
We interpret this as a very clear indication that our scheme succeeds to extract large resummable contributions in the perturbative region, which is overlapped with non-perturbative model within the CSS approach.

\section{Definition of Quark-TMDPDF}

Extending the work done in~\cite{Echevarria:2012js,GarciaEchevarria:2011rb}, we define in impact parameter~\footnote{All quantities with tilde in this paper refer to quantities calculated in impact parameter space.} space a quark-TMDPDF of a polarized hadron, collinear in the $+z$ direction with momentum $P$ and spin $\vec S$ as
\begin{align}\label{eq:tmddef}
\tilde F_{n,\a\b} &=
\tilde{\Phi}_{n,\a\b}^{(0)}(\D)\,
\sqrt{\tilde{S}\le(\D,\D\ri)}\,,
\end{align}
where we have used the $\D$-regulator as a particular choice to regulate the rapidity divergencies.
$\Phi_{n,\a\b}^{(0)}$ stands for a purely collinear matrix element, i.e., a matrix element which has no overlap with the soft region~\cite{Manohar:2006nz}, and it is given by the bilocal correlator
\begin{align}
\Phi_{n,\a\b}^{(0)}
&=
\langle P\vec S|\,\le[\bar\x_{n\a} W^T_n\ri](0^+,y^-,\vec y_\perp)\,
\le[W_n^{T\dagger} \x_{n\b}\ri](0)\,| P\vec S\rangle
\,.
\end{align}
The soft function $S$, which encodes soft-gluon emission (for more details about the relevance of its contribution the reader can consult Ref.~\cite{GarciaEchevarria:2011rb}), is given by
\begin{align}
S &=
\sandwich{0}{ {\rm Tr} \; \Big[S_n^{T\dagger} S^T_\bn \Big](0^+,0^-,\vec y_\perp)\le[S^{T\dagger}_\bn S^T_n\ri](0)}{0}\,.
\end{align}
We should mention that $\tilde F_{n,\a\b}$ is free from all rapidity divergences, which cancel in the combination of the collinear and soft matrix elements in Eq.~(\ref{eq:tmddef}), and thus the only $\D$-dependence that it contains is pure infrared~\cite{Echevarria:2012js}.

To obtain the eight leading-twist quark-TMDPDFs~\cite{Mulders:1995dh,Boer:1997nt}, represented generically by $\tilde{F_{n}}$ below, one can simply take the trace of $\tilde{F}_{n,\a\b}$ with the Dirac structures $\frac{\bnslash}{2}$, $\frac{\bnslash \g_5}{2}$ and $\frac{i\s^{j+}\g_5}{2}$ for
unpolarized, longitudinally polarized and transversely polarized quarks, respectively, inside a polarized hadron.
The superscript $T$ indicates transverse gauge-links $T_{n(\bn)}$, necessary to render the matrix elements gauge-invariant~\cite{Idilbi:2010im,GarciaEchevarria:2011md}. The definitions of collinear ($W_{n(\bn)}$), soft ($S_{n(\bn)}$) and transverse
($T_{n(\bn)}$) Wilson lines for DY and DIS kinematics can be found in~\cite{GarciaEchevarria:2011rb}.

In~\cite{GarciaEchevarria:2011rb} the anomalous dimension of the unpolarized TMDPDF was given up to  3-loop order based on
a factorization theorem for $q_T$-dependent observables in a Drell-Yan process.
Such a factorization theorem for the hadronic tensor can be written in impact parameter space, using the definition of the TMDPDF given in Eq.~(\ref{eq:tmddef}), as
\begin{align}\label{eq:factth}
\tilde{M} &=
H(Q^2/\m^2)\,
\tilde{F}_{n}(x_n,b;Q,\m)\,
\tilde{F}_{\bn}(x_\bn,b;Q,\m)
+{\cal O}\le( \le(bQ\ri)^{-1}\ri)
\,.
\end{align}
where $\m$ stands for renormalization/factorization scale that separates the perturbative and non-perturbative physics, as it is customary in a quantum field theory like QCD. 
$H$ is the hard coefficient encoding the physics at the probing scale $Q$ and which is a polynomial of only $\ln(Q^2/\mu^2)$.
This quantity is built, to all orders in perturbation theory, by considering virtual Feynman diagrams only, and no real gluon emission has to be considered (even in diagrams with mixed real and gluon contributions). Moreover, the quantity $H$ has to be free from infrared physics, no matter how the latter is regularized. This is a general principle and it should work whether one works on or off-the-light-cone.

Since the factorization theorem given above holds, at leading-twist, also for spin-dependent observables, one can apply the same arguments as for the unpolarized case, based on renormalization group invariance, to get a relation between the anomalous dimensions of $\tilde F$ and $H$.
Since the anomalous dimensions of the two TMDPDFs in Eq.~(\ref{eq:factth}) are identical \cite{Echevarria:2012js}, we have
\begin{align}
\g_F &=
-\frac{1}{2}\g_H =
-\frac{1}{2} \le[ 2\G_{\rm cusp}\, \ln\frac{Q^2}{\m^2} + 2\g^V \ri]
\,,
\end{align}
where $\g_H$ is known at 3-loop level~\cite{Idilbi:2006dg,Moch:2005id,Moch:2004pa} (see Appendix \ref{sec:app} for more details).
$\G_{\rm cusp}$ stands for the well-known cusp anomalous dimension in the fundamental representation.
This crucial result can be automatically extended to the eight leading-twist quark-TMDPDFs defined in Eq.~(\ref{eq:tmddef}), since the anomalous dimension is
independent of spin structure.

\section{Evolution Kernel}

For spin-dependent TMDPDFs the OPE in terms of collinear PDFs fails.
For instance, the Sivers function at large $q_T$ is matched onto a twist-3 collinear operator~\cite{Aybat:2011ge}.
Since the phenomenological extraction of TMDs is much more difficult than of integrated PDFs, one has to resort to non-perturbative models as the starting point for scale evolution and fit their parameters by comparison to data.
Obviously, knowing the evolution of these hadronic matrix elements to the highest possible accuracy is very beneficial.

Starting from Eq.~(\ref{eq:tmddef}) the evolution of a generic quark-TMDPDF between initial factorization scale $\m_i$ and probing scale $Q_i$ and final ones $\m_f$ and $Q_f$ is given by~\footnote{Since the evolution kernel is the same for $\tilde{F}_{n}$ and $\tilde{F}_{\bn}$, we have dropped out the $n,\bn$ labels.}
\begin{align}\label{eq:tmdevolution}
\tilde F(x,b;Q_f,\m_f)& = \tilde F(x,b;Q_i,\m_i)\, \tilde R(b;Q_i,\m_i,Q_f,\m_f)\,,
\end{align}
where the evolution kernel $\tilde R$ is~\cite{GarciaEchevarria:2011rb,Echevarria:2012js}
\begin{align}\label{eq:tmdkernel}
\tilde R(b;Q_i,\m_i,Q_f,\m_f) &=
\exp\le\{
\int_{\m_i}^{\m_f} \frac{d\bar\m}{\bar\m} \g_F\le(\as(\bar\m),\ln\frac{Q_f^2}{\bar\m^2} \ri)
\ri\}
\le( \frac{Q_f^2}{Q_i^2} \ri)^{-D\le(b;\m_i\ri)}
\,.
\end{align}

As explained in~\cite{Echevarria:2012js}, this evolution kernel is identical to the one that can be extracted from Collins' approach to TMDs~\cite{Collins:2011zzd} when one identifies $\sqrt{\z_i}=Q_i$ and $\sqrt{\z_f}=Q_f$.
Moreover, below we will choose $\m_i=Q_i$ and $\m_f=Q_f$ to illustrate the application of the kernel.

The $D$ term can be obtained by noticing that the renormalized $\tilde{F}$ has to be well-defined when its partonic version is calculated pertubatively. This means that all divergences, other than genuine long-distance ones, have to cancel. This fundamental statement, that rapidity divergences cancel when the collinear and soft matrix elements are combined according to Eq.~(\ref{eq:tmddef}), allows one to
extract all the $Q^2$-dependence from the TMDPDFs and exponentiate it with the $D$ term (see Sec.~5 in~\cite{GarciaEchevarria:2011rb}),
thereby summing large logarithms $\ln(Q^2/q^2_T)$.
Applying renormalization group invariance to the hadronic tensor $\tilde{M}$ in~Eq.~(\ref{eq:factth}) we get the following relation,
\begin{align}\label{eq:drg}
\frac{dD}{d\ln\m} &= \G_{\rm cusp}\,,
\end{align}
where the cusp anomalous dimension $\G_{\rm cusp}$ is known at three-loops~\cite{Moch:2004pa}.

In the case of semi-inclusive deep inelastic scattering (SIDIS) or the $q_T$-spectrum of Higgs boson production, for instance, the relevant factorization theorems are analogous to the one given in Eq.~(\ref{eq:factth}) for Drell-Yan production. 
Essentially, they consist of a hard part and two TMDs, either a quark-TMDPDF and a quark-TMD fragmentation function (TMDFF) for SIDIS or two gluon-TMDPDFs for Higgs boson production.
Therefor, apart from the necessary changes that have to be made to $\g_F$ and $\G_{\rm cusp}$ while considering quark/gluon-TMDPDF or quark/gluon TMDFF, the form of the evolution kernel for all TMDs is similar to the one given in Eq.~(\ref{eq:tmdkernel}).
Then, below we will generally refer to TMDs, since it is straightforward to apply the machinery developed in this work to different kinds of TMDs.

The evolution of TMDs, given by the evolution kernel in Eq.~(\ref{eq:tmdkernel}), is done in impact parameter space, thus we need to Fourier transform back to momentum space and large logarithms $L_\perp=\ln(\m_i^2 b^2 e^{2\g_E}/4)$ will appear then in the $D(b;\m_i)$ term when $b$ is either large or small.
We resum these logarithms in order to get numerically more precise predictions.

The resummation that we present in the next section is valid only within the perturbative domain of the impact parameter, $b\lesssim{\cal O}(1/\lqcd)$. Outside this region we need a non-perturbative model for the $D(b;\m_i)$ term.
However our aim is to characterize the perturbative region and obtain a parameter free expression for the kernel using all the existing information on $\G_{\rm cusp}$ and fixed order calculations of the $D$ term.
Under certain circumstances, as we will show, we find that knowing the evolution kernel only in its perturbative domain is enough to evolve the TMDs in a model independent way.
As a result, all the model dependence will be restricted to the functional form of the low energy TMDs to be extracted from fitting to data.

Below we provide the resummation of large $L_\perp$ logarithms based on the spirit of effective field theories, i.e., leaving fixed the scale within the strong coupling constant.
Instead of solving directly the renormalization group evolution in Eq.~(\ref{eq:drg}) as it is done within the standard CSS approach, we derive a recursive relation for the coefficients of the perturbative expansion of the $D$ term and solve it to resum the large logarithms to all orders.

\subsection{Derivation of $D^R$}

Matching the perturbative expansions of the $D$ term,
\begin{align}\label{eq:dseries}
D(b;\m_i) = \sum_{n=1}^\infty d_n(L_\perp) \left( \frac{\as(\m_i)}{4\pi} \right)^n\,,
\quad\quad
L_\perp=\ln\frac{\m_i^2 b^2}{4e^{-2\g_E}}
\,,
\end{align}
the cusp anomalous dimension $\G_{\rm cusp}$ and the QCD $\b$-function (see Appendix~\ref{sec:app}), one gets the following recursive differential equation
\begin{align}\label{eq:ddif}
d_n^\prime(L_\perp)&=\frac{1}{2}\Gamma_{n-1}+\sum_{m=1}^{n-1}m \beta_{n-1-m} d_m(L_\perp)\,,
\end{align}
where $d'_n\equiv d d_n/dL_\perp$.
Solving this equation one can get the structure of the first three $d_n$ coefficients
\begin{align}\label{eq:dcoeffs}
d_1(L_\perp) &=
\frac{\G_0}{2 \b_0} \le(\b_0 L_\perp \ri)
+ d_1(0)
\,,
\nn\\
d_2(L_\perp) &=
\frac{\G_0}{4\b_0} \le(\b_0 L_\perp \ri)^2
+ \le(\frac{\G_1}{2\b_0} + d_1(0)\ri) \le(\b_0 L_\perp \ri)
+ d_2(0)
\,,
\nn\\
d_3(L_\perp) &=
\frac{\G_0}{6\b_0} \le(\b_0 L_\perp \ri)^3
+ \frac{1}{2}\le(\frac{\G_1}{\b_0} + \frac{1}{2}\frac{\G_0\b_1}{\b_0^2} + 2d_1(0) \ri)
\le(\b_0 L_\perp \ri)^2
+
\frac{1}{2} \le( 4d_2(0) + \frac{\b_1}{\b_0} 2d_1(0)
+ \frac{\G_2}{\b_0}\ri) \le(\b_0 L_\perp \ri)
+ d_3(0)
\,.
\end{align}
From known perturbative calculations of the Drell-Yan cross section we can fix the first two finite coefficients, as is explained in~\cite{Becher:2010tm}~\footnote{In the notation of~\cite{Becher:2010tm}, our $d_n(0)$ corresponds to their $d_n^q/2$.}:
\begin{align}
d_1(0) &= 0
\,,
\nn\\
d_2(0) &=
C_F C_A \left(\frac{404}{27}-14\z_3\right)
- \left(\frac{112}{27}\right)C_F T_F n_f
\,.
\end{align}
Now, based on the generalization of Eq.~(\ref{eq:dcoeffs}) for any value of $n$ and after some tedious algebra, we can derive the general form of the $d_n(L_\perp)$ coefficients, being
\begin{align}\label{eq:2dn}
2 d_n(L_\perp) &=
(\beta_0 L_\perp)^n \le(\frac{\Gamma_0}{\beta_0}\frac{1}{n}\ri)
+ (\beta_0 L_\perp)^{n-1} \le(
\frac{\Gamma_0\beta_1}{\beta_0^2}\le(-1+H^{(1)}_{n-1}\ri)|_{n\geq 3}
+ \frac{\Gamma_1}{\beta_0}|_{n\geq 2}
\ri)
\nn\\
&+
(\beta_0 L_\perp)^{n-2} \le( (n-1)2d_2(0)|_{n\geq 2}
+ (n-1)\frac{\Gamma_2}{2\beta_0}|_{n\geq 3}
+ \frac{\beta_1\Gamma_1}{\beta_0^2}s_n|_{n\geq 4}
+\frac{\beta_1^2\Gamma_0}{\beta_0^3}t_n|_{n\geq 5}
+\frac{\beta_2\Gamma_0}{2\beta_0^2} (n-3)|_{n\geq 4}\right) +...\,,
\end{align}
where
\begin{align}
s_n &=
(n-1) H^{(1)}_{n-2}+\frac{1}{2} (5-3 n)\,,
\nn\\
t_n &=
\frac{1}{2} \left[(1-n) H_{n-1}^{(2)}+ (n+1)
+(n-1) (\psi(n)+\g_E -2) (\psi(n)+\g_E )\right]\,.
\end{align}
$H_{n}^{(r)}=\sum_{m=0}^{n} m^r$ is the $r$-th order Harmonic Number function of $n$ and
$\psi(n)=\G'(n)/\G(n)$ is the digamma function of $n$.
Setting $a=\as(\m_i)/(4\pi)$, $X=a\beta_0 L_\perp$ and using the previous result we write the $D$ term as
\begin{align}
D^R(b;\m_i) &= \sum_{n=1}^{\infty} d_n(L_\perp)  a^n =
\nn\\
&\frac{1}{2}\sum_{n=1}^{\infty} \le\{
X^n \le(\frac{\Gamma_0}{\beta_0}\frac{1}{n} \ri)
+ a X^{n-1} \le(
\frac{\Gamma_0\beta_1}{\beta_0^2}\le(-1+H^{(1)}_{n-1}\ri)|_{n\geq 3}
+ \frac{\Gamma_1}{\beta_0}|_{n\geq 2}
\ri)
\ri.
\nn\\
&\le.
+ a^2 X^{n-2}
\le( (n-1)2 d_2(0)|_{n\geq 2}
+ (n-1)\frac{\Gamma_2}{2\beta_0}|_{n\geq 3}
+ \frac{\beta_1\Gamma_1}{\beta_0^2}s_n|_{n\geq 4}
+\frac{\beta_1^2\Gamma_0}{\beta_0^3}t_n|_{n\geq 5}
+\frac{\beta_2\Gamma_0}{2\beta_0^2} (n-3)|_{n\geq 4}\right) +...
\ri\}
\,,
\end{align}
where the label $R$ stands for ``resummed''.
Once we have the series of the $D$ term organized as above, each order in $a$ can be summed for $|X|<1$, giving
\begin{align}\label{eq:resummedD}
 D^R(b;\m_i) &=
-\frac{\Gamma_0}{2\beta_0}\ln(1-X)
+ \frac{1}{2}\le(\frac{a}{1-X}\ri) \le[
- \frac{\beta_1\Gamma_0}{\beta_0^2} (X+\ln(1-X))
+\frac{\Gamma_1}{\beta_0} X\ri]
\nn\\
&+ \frac{1}{2}
\le(\frac{a}{1-X}\ri)^2\le[
2d_2(0)
+\frac{\Gamma_2}{2\beta_0}(X (2-X
))
+\frac{\beta_1\Gamma_1}{2 \beta_0^2} \le( X (X-2)-2 \ln (1-X)\ri)
+\frac{\beta_2\Gamma_0}{2\beta_0^2} X^2 
\ri.
\nn\\
&\le.
+\frac{\beta_1^2\Gamma_0}{2 \beta_0^3} (\ln^2(1-X)-X^2)
\ri]
 +...
\,,
\end{align}

As is clear from Eq.~(\ref{eq:resummedD}) this result for $D^R$ can be analytically continued through Borel-summation and its validity can thus be extended to $X\to-\infty$, which corresponds to $b\to 0$ (see Eq.~(\ref{eq:dseries})).
The maximum value of $X$ where each coefficient of $a^n$ in Eq.~(\ref{eq:resummedD}) is valid is $X=1$.
This upper limit corresponds to
\begin{align}
b_X(\m_i) &=
\frac{2e^{-\g_E}}{\m_i}\exp\le[\frac{2\pi}{\b_0 \as(\m_i)}\ri]
\,.
\end{align}
In Appendix~\ref{sec:uyy} we provide also the expression of $D^R$ up to NNNLL. 
At that order one would need the cusp anomalous dimension at ${\cal O}(\alpha^4_s)$ and $d_3(0)$, which presently are not known.

\subsection{Range of Validity of $D^R$ and the Landau Pole}

\begin{figure}[t]
\begin{center}
\includegraphics[width=0.45\textwidth]{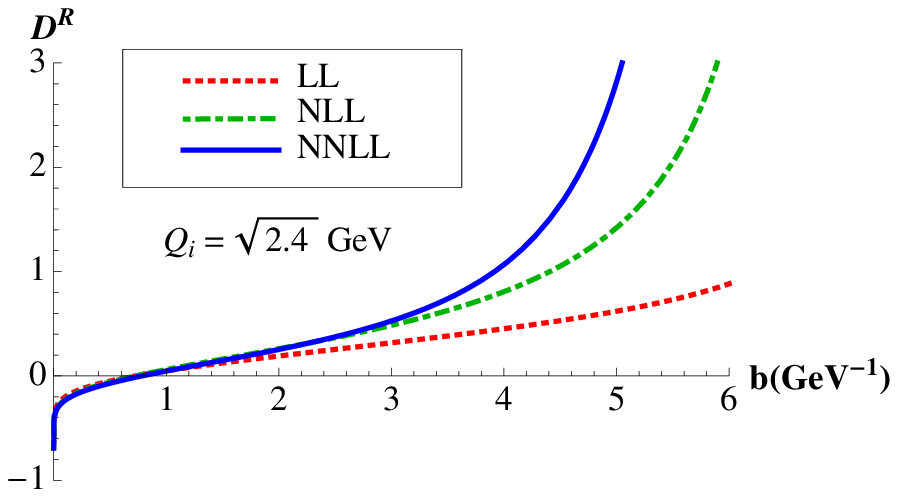}
\quad\quad\quad
\includegraphics[width=0.45\textwidth]{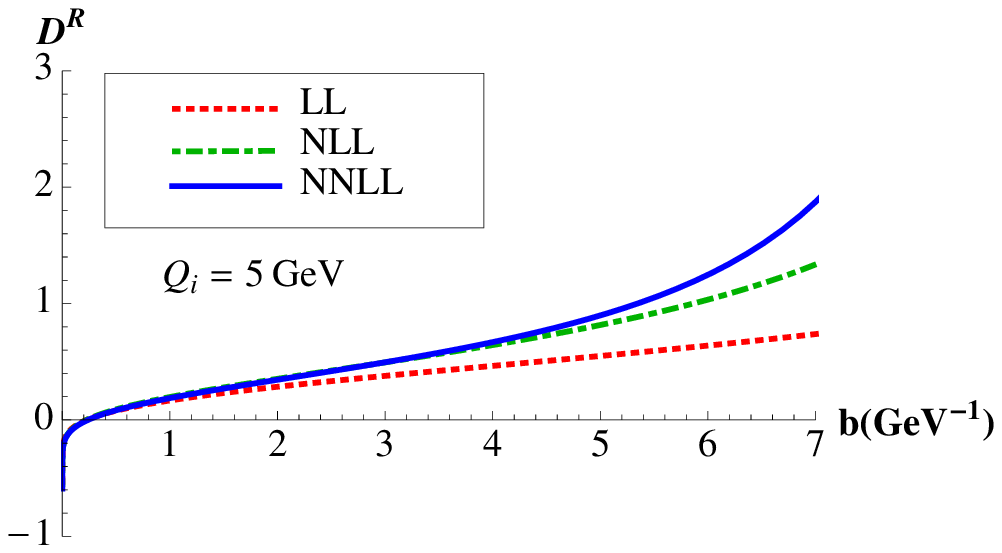}
\\
\hspace{0cm}(a)\hspace{9cm}(b)\hspace{6cm}
\end{center}
\caption{\it
Resummed D at $Q_i=\sqrt{2.4}~{\rm GeV}$ with $n_f=4$ (a) and $Q_i=5~{\rm GeV}$ with $n_f=5$ (b).
}
\label{fig:resummedD}
\end{figure}

Although each order in $a$ in Eq.~(\ref{eq:resummedD}) is valid for $0<b<b_X$, the interval of convergence of the series is smaller.
The fact that each term diverges at $b_X$ makes the series itself more and more divergent as we approach this point. 
In Fig.~\ref{fig:resummedD} we show the $D^R$ for two different scales, from which it is clear that the convergence between leading logarithm (LL), next-to leading logarithm (NLL) and next-to-next-to leading logarithm (NNLL) is extremely good for small values of $b$ and gets spoiled as we approach $b_X$.
It is also evident from Fig.~\ref{fig:resummedD} how the range of convergence changes as we vary the initial scale $\m_i$, since $b_X$ depends on this scale.
It is interesting then to study the behavior of $D^R$ analytically when the impact parameter approaches $b_X$, which is the kinematic region where the analysis becomes more subtle.

The fact that the convergence of $D^R$ gets spoiled around $b_X$ is related to the Landau pole. 
Although the scale in the strong coupling is fixed, $\as(\m_i)$, the effects of non-perturbative physics are ``shifted'' to the coefficients of the perturbative expansion of the $D^R$ term, which grow and ultimately lead to the breakdown of the perturbative series.
Thus, in our approach the issue of the Landau pole reemerges as the divergence at $X=1$.
In fact, using the usual expansion of $\Lambda_{\rm QCD}=Q \exp\le[G(t_{Q})\ri]$, where $t_{Q}\equiv -2\pi/(\beta_0\alpha_s(Q))$ and
\begin{align}\label{eq:gt}
G(t) &=
t + \frac{\beta_1}{2 \beta^2_0}\ln(-t)
- \frac{\beta_1^2-\beta_0\beta_2}{4\beta_0^4}\frac{1}{t}
- \frac{\beta_1^3-2\beta_0\beta_1\beta_2+\beta_0^2\beta_3}{8\beta_0^6}\frac{1}{2t^2} + \dots
\,,
\end{align}
we have
\begin{align}
b_X &= A(\m_i)\,b_{\L_{\rm QCD}}
\,,\quad\quad
A(\m_i) = \exp(-t_{\m_i}+G(t_{\m_i}))
\,, \quad\quad
b_{\L_{\rm QCD}} =
\frac{2 e^{-\gamma_E}}{\Lambda_{\rm QCD}}
\,,
\end{align}
from which it is clear that $b_X$ is closely related to $b_{\L_{\rm QCD}}$, up to the $\m_i$-dependent proportionality factor $A(\m_i)$.
Given Eq.~(\ref{eq:gt}), at LL accuracy $G(t)=t$, and thus $A(\m_i)=1$ and $b_X=b_{\lqcd}$ at that accuracy.
When one goes beyond LL accuracy for $G(t)$ and considers the available information on the $\b$-function as illustrated in Eq.~(\ref{eq:gt}), numerically one finds that $1\leq A(\m_i)\leq 2$  for 1~GeV$\leq \m_i\leq$ 1~TeV.
We conclude that the divergence of $D^R$ at $X=1$ is a manifestation of the Landau pole, as claimed before.

One can calculate the numerical value of $\lqcd$, which for $n_f=5$ and $\alpha_s(M_Z)=0.117$ is $\lqcd \approx 157~{\rm MeV}$, and correspondingly $b_{\lqcd} \approx 7~{\rm GeV}^{-1}$.
At this point we are clearly within the non-perturbative region, which cannot be accessed by perturbative calculations and has to be modeled and extracted from experimental data.

In Section~\ref{sec:DRall}  and in Appendix~\ref{sec:uyy} we show how to derive an expression for $D^R$ up to any desired perturbative order. 
Using Eqs.~(\ref{eq:resummedD}) and~(\ref{eq:C4}) we get the asymptotic expression of $D^R$ when $X\sim 1$ at NNNLL,
\begin{align}\label{eq:DRasy}
D^R |_{X\to 1^-}  &=
-\frac{\G_0}{2\b_0}\ln(1-X) \le[
1 + \le(\frac{a}{1-X} \ri) \frac{\b_1}{\b_0}
+ \le(\frac{a}{1-X} \ri)^2 \frac{\beta_1}{\beta_0}\left(\frac{\G_1}{\G_0}-\frac{\b_1}{\b_0}\ln(1-X)\right)
\ri. \nn \\ &
\le.
 +
\le(\frac{a}{1-X} \ri)^3 \frac{\beta_1}{\beta_0}
\left(\frac{\beta_1^2}{3\beta_0^2}\ln^2(1-X)
-\left(\frac{\G_1\b_1}{\G_0\b_0}+\frac{\b_1^2}{2\b_0^2}
\right)
\ln(1-X)+
\frac{\Gamma_2}{\Gamma_0}+\frac{\b_2}{\b_0}-\frac{\b_1^2}{\b_0^2}
\ri)
+ ...
\ri]
\,,
\end{align}
from which one can obtain (approximately) the values of $b$ where convergence is lost. This can also be inferred from Fig.~\ref{fig:resummedD}.
Thus we can trust $D^R$ up to $b_c\sim 4~{\rm GeV}^{-1}$ for $\m_i=\sqrt{2.4}~{\rm GeV}$ and $b_c\sim 6~{\rm GeV}^{-1}$ for $\m_i=5~{\rm GeV}$.
Notice that we have used different numbers of active flavors depending on the scale $\m_i$, $n_f=4$ for $\m_i=\sqrt{2.4}~{\rm GeV}$ and $n_f=5$ for $\m_i=5~{\rm GeV}$, since we have set the threshold of the bottom mass to $m_b=4.2~{\rm GeV}$.
It is clear then that the larger the initial scale $\m_i$ is the broader the interval of the impact parameter where the convergence of $D^R$ is acceptable, and where $b_{\lqcd}$ is the maximum achievable value.
The two cases shown in Fig.~\ref{fig:resummedD}  represent two extreme phenomenological cases, between which one should choose the initial scale in order to fix the low energy models for TMDs. 

A last comment worth mentioning concerns the convergence of $D^R$ in the small $b$ region. 
As discussed above, the convergence of the resummed $D$ is only spoiled in the region around the Landau pole, i.e., for $b$ close to $b_{\Lambda_{\rm QCD}}$. 
In the small $b$ region $D^R$ is completely resummable (see Fig.~\ref{fig:resummedD}) and this agrees with other studies on the perturbative series in this region~\cite{Becher:2010tm}. 

Summarizing, the resummation method explained above allows us to implement the evolution kernel just in a finite range of the impact parameter while for larger values of $b$ one clearly needs a non-perturbative contribution. 
The discussion of such contribution is beyond the scope of the current work.
Then, we can write
\begin{align}\label{eq:Rcnp}
\tilde R(b;Q_i,\m_i,Q_f,\m_f) &=
\exp\le\{
\int_{\m_i}^{\m_f} \frac{d\bar\m}{\bar\m} \g_F\le(\as(\bar\m),\ln\frac{Q_f^2}{\bar\m^2} \ri)
\ri\}
\le( \frac{Q_f^2}{Q_i^2} \ri)^
{-\le[D^R\le(b;\m_i\ri)\theta(b_c-b)+D^{NP}(b;\m_i)\theta(b-b_c)\ri]}
\,,
\end{align}
where $D^{NP}$ stands for the non-perturbative piece of the $D$ term and $b_c<b_{\L_{\rm QCD}}$ depends on the scale $\m_i$ as explained before.

\subsection{Applicability of the Evolution Kernel}

\begin{figure}[t]
\begin{center}
\includegraphics[width=0.45\textwidth]{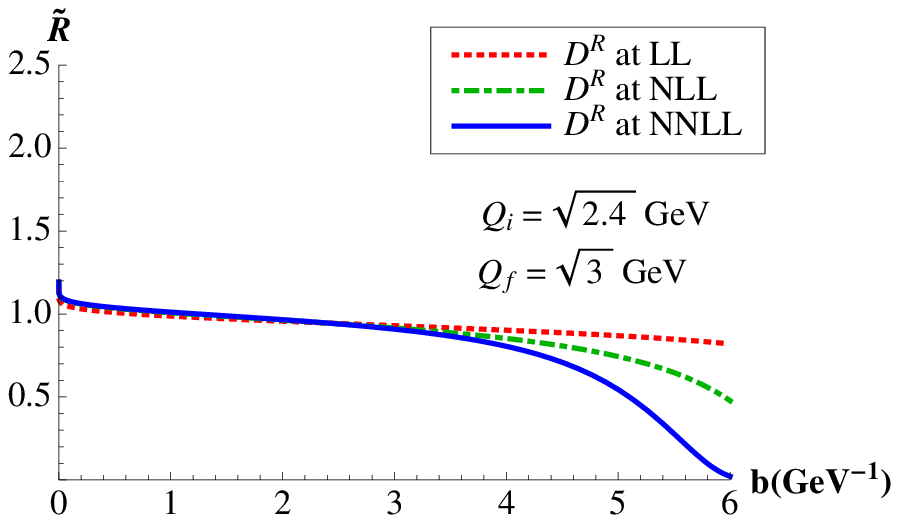}
\quad\quad\quad
\includegraphics[width=0.45\textwidth]{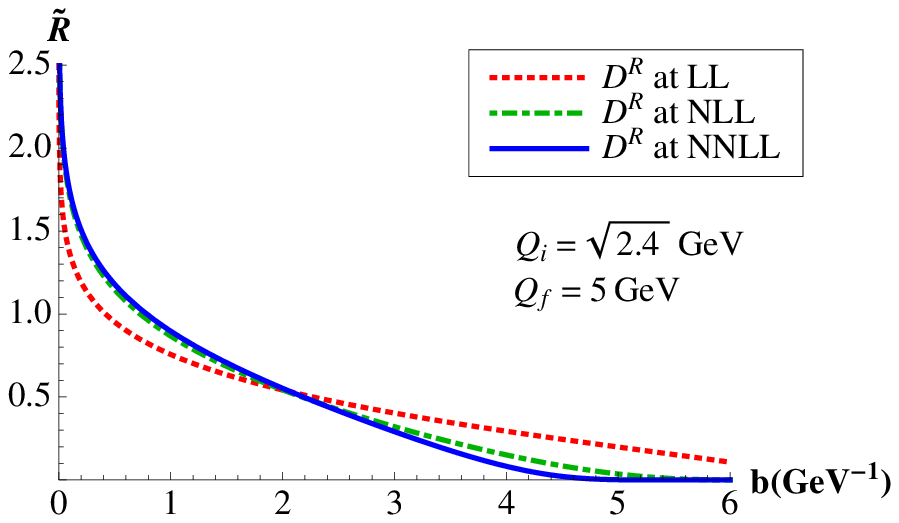}
\\
\hspace{0cm}(a)\hspace{9cm}(b)\hspace{6cm}
\\
\includegraphics[width=0.45\textwidth]{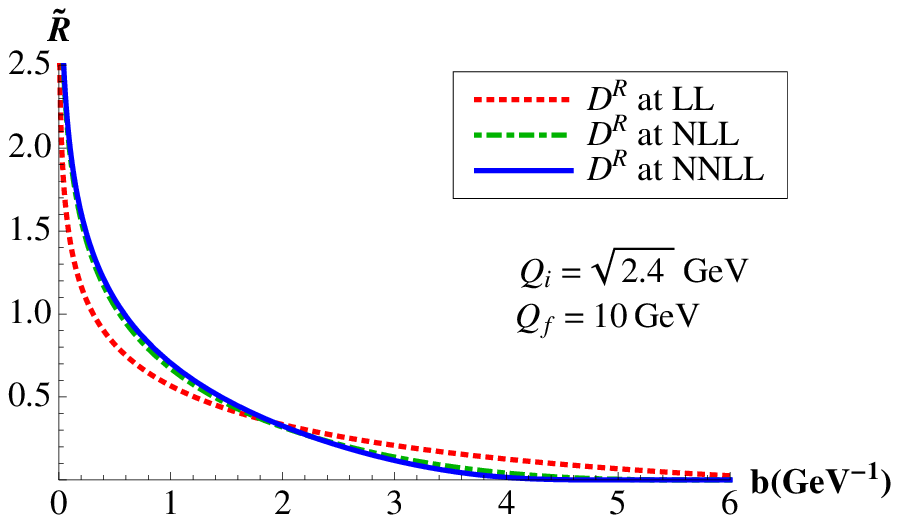}
\quad\quad\quad
\includegraphics[width=0.45\textwidth]{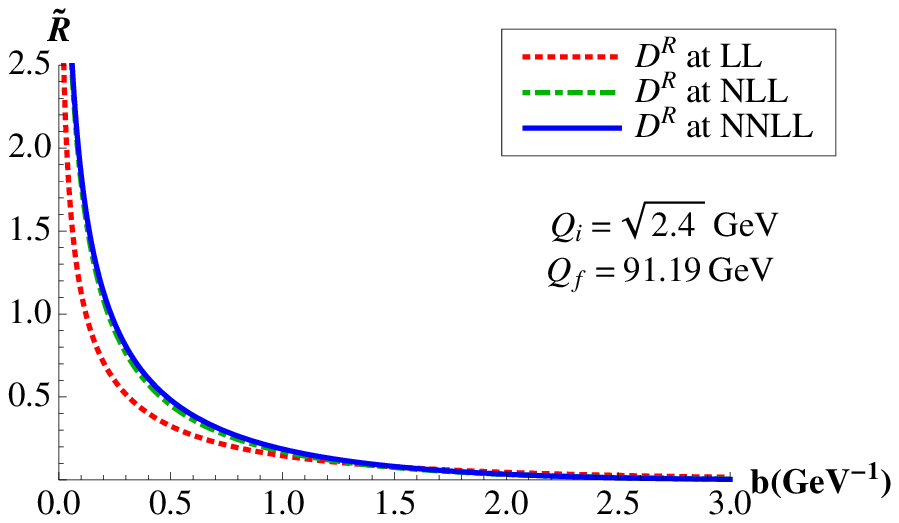}
\\
\hspace{0cm}(c)\hspace{9cm}(d)\hspace{6cm}
\end{center}
\caption{\it
Evolution kernel from $Q_i=\sqrt{2.4}~{\rm GeV}$ up to $Q_f=\{\sqrt{3}\,,5\,,10,91.19\}~{\rm GeV}$.
}
\label{fig:kernelDR}
\end{figure}

As explained in the previous section, we can obtain perturbatively the evolution kernel only in a finite range of the impact parameter.
Since our aim is to reduce as much as possible the need to introduce models (and free parameters) for the evolution kernel itself and to shift all such model dependence to the input low-energy TMDs, we need to find under which conditions the effects of the large $b$ region are suppressed when evolving the TMDs.
Choosing $\m_i=Q_i$ and $\m_f=Q_f$ to simplify the discussion, our goal is to be able to apply the following expression,
\begin{align}\label{eq:Rc}
\tilde R(b;Q_i,Q_f) &=
\exp\le\{
\int_{Q_i}^{Q_f} \frac{d\bar\m}{\bar\m} \g_F\le(\as(\bar\m),\ln\frac{Q_f^2}{\bar\m^2} \ri)
\ri\}
\le( \frac{Q_f^2}{Q_i^2} \ri)^
{-D^R\le(b;Q_i\ri)} \theta(b_c-b)
\,,
\end{align}
which is parameter free.
In order to fix when this is the case, one needs to consider both the range in which $D^R$ converges, as shown in Fig.~\ref{fig:resummedD}, and the final scale $Q_f$ up to which we are evolving the TMDs, as shown in Fig.~\ref{fig:kernelDR}.

To start with, we already showed that the range of convergence of $D^R$ depends on the initial scale $Q_i$.
However we are interested in cases where the final scale $Q_f$ is large enough so that the kernel itself becomes vanishingly small in the large $b$ region. 
In fact, the evolution kernel in Eq.~(\ref{eq:Rc}) is actually the exponential of $-D^R\ln(Q_f^2/Q_i^2)$, which guarantees that when $b \to b_X^-$ ($X\to 1^-$), one has $\tilde R\to 0$ for $Q_f>Q_i$, due to the sign of the exponent.
For the leading order term in Eq.~(\ref{eq:resummedD}) we have
\begin{align}
\lim_{b\to b_X^-} D^R_0 &=
\lim_{b\to b_X^-}\le[
-\frac{\G_0}{2\b_0} \ln(1-X) \ri] \to
+ \infty
\,,
\end{align}
and this limit is not spoiled by higher order corrections, as we show in Figs.~\ref{fig:resummedD} and~\ref{fig:kernelDR}.
Thus, the larger $Q_f$ is compared to $Q_i$, the faster the kernel goes to zero, as it is clear from Fig.~\ref{fig:kernelDR}. 
The fact that we have at our disposal several perturbative orders is essential to test the convergence of the evolution kernel and of the evolved TMDs.

In Fig.~\ref{fig:kernelDR}a we show the evolution kernel in Eq.~(\ref{eq:tmdkernel}) with $D^R$ given in Eq.~(\ref{eq:resummedD}) for a final scale $Q_f$ which is quite close to the initial one $Q_i$.
In this case we would expect the kernel to be nearly $1$ and to decrease smoothly for large values of $b$, since the TMD to be evolved is not supposed to change dramatically its shape.
However, we see that the convergence of the kernel fails around $b_c\sim 4~{\rm GeV}^{-1}$, which is consistent with Fig.~\ref{fig:resummedD}.
As already explained, we can only trust the perturbative implementation of the kernel up to $b_c$, where the $D^R$ starts to diverge, and clearly if $Q_f$ is not large enough, Eq.~(\ref{eq:Rc}) does not give us a proper approximation for the kernel.
On the other hand, we can see from Figs.~\ref{fig:kernelDR}b, \ref{fig:kernelDR}c and \ref{fig:kernelDR}d that the larger $Q_f$ is, the faster the kernel decreases.
And although we cannot access the kernel in the non-perturbative region, in the particular case where $Q_i=\sqrt{2.4}~{\rm GeV}$ and $Q_f\gtrsim 5~{\rm GeV}$, it is negligible for $b\gtrsim b_c$.
Simply stated, we look for the kinematical setup where the non-perturbative region of the kernel is suppressed.
This allows us to implement safely Eq.~(\ref{eq:Rc}), which gives us a very good approximation and a parameter free kernel for all practical purposes.
Using this kernel within the already explained kinematical setup, i.e., as long as $Q_f$ is large enough compared to $Q_i$, we can evolve low energy models for TMDs and extract their parameters by fitting to data.
The advantage in this case is that all the model dependence is restricted to the functional form of TMDs, while the evolution itself is implemented perturbatively, i.e., without any free parameters.

\section{An Alternative Extraction of $D^R$}
\label{sec:DRall}

\begin{figure}[t]
\begin{center}
\includegraphics[width=0.45\textwidth]{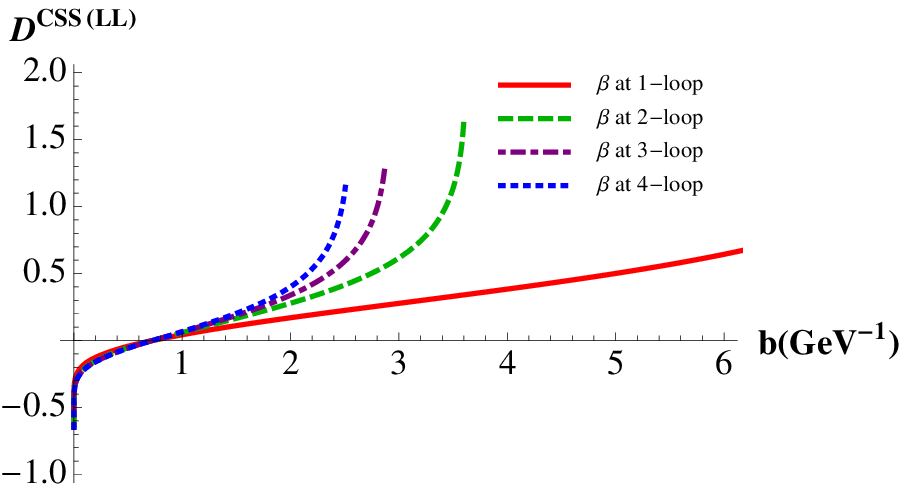}
\quad\quad\quad
\includegraphics[width=0.45\textwidth]{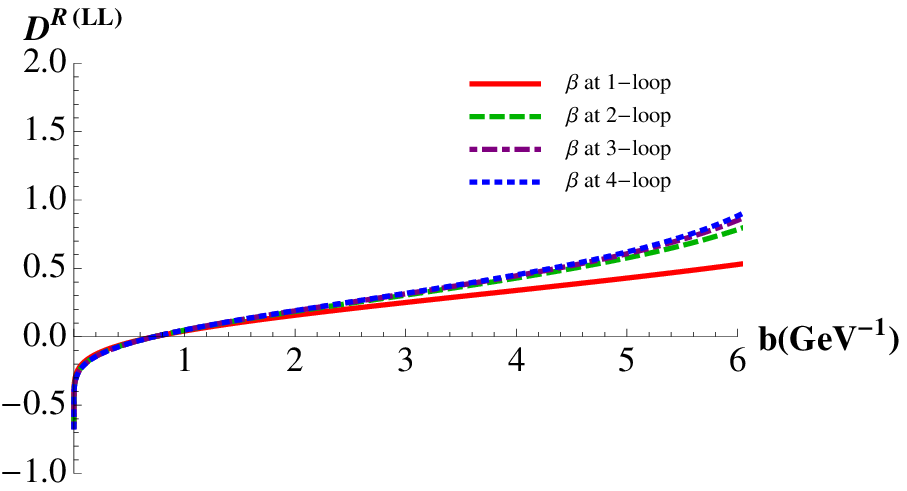}
\end{center}
\caption{\it
Resummed $D(b;Q_i=\sqrt{2.4})$ at LL of Eqs.~(\ref{eq:CSSD1}), (a), and~(\ref{eq:CSSD2}), (b), with the running of the strong coupling  at various orders and decoupling coefficients included.
}
\label{fig:DLLorders}
\end{figure}

The expression for $D^R$,  Eq.~(\ref{eq:resummedD}), was obtained  by solving a recursive differential equation that resulted from the RG-evolution of the $D$ in Eq.~(\ref{eq:drg}).
Below we show that one can derive the same result  by solving a differential equation as it is done within the standard CSS approach, i.e.,
\begin{align}\label{eq:devolved0}
D\le(b;Q_i\ri) &=
D\le(b;\m_b\ri)
+ \int_{\m_b}^{Q_i}\frac{d\bar\m}{\bar\m} \G_{\rm cusp}
\,,
\end{align}
where $\mu_b=2e^{-\g_E}/b$ to cancel the $L_\perp$ logarithms.
For this it will be necessary to consider the running of the strong coupling properly with the resummation scheme.

First, we integrate Eq.~(\ref{eq:devolved0}), getting at lowest order in perturbation theory,
\begin{align}
\label{eq:CSSD1}
D(b; Q_i)&=-\frac{\Gamma_0}{ 2 \beta_0}\ln\frac{\alpha_s(Q_i)}{\alpha_s(\mu_b)}
\,.
\end{align}
Re-expressing $\alpha_s(\mu_b)$ in terms of $\alpha_s(Q_i)$  at the  correct perturbative order, $\alpha_s(\mu_b)=\alpha_s(Q)/(1-X)$, one finds
\begin{align}
\label{eq:CSSD2}
D(b;Q_i)&=-\frac{\Gamma_0}{ 2 \beta_0}\ln (1-X) \ ,
\end{align}
which coincides with the first term of the r.h.s of Eq.~(\ref{eq:resummedD}).
Repeating the same steps at higher orders one gets that the resummed  $D$ within CSS approach given in Eq.~(\ref{eq:devolved0}) and our case, given in Eq.~(\ref{eq:resummedD}), are exactly the same order by order.
In Appendix~\ref{sec:uyy} we give an explicit derivation of the same result at NLL, NNLL and NNNLL as well.
The expansion of the $D$ as in Eq.~(\ref{eq:CSSD1}) at NLL, NNLL and NNNLL is given in Eq.~(\ref{eq:C2}). 
Thus, we conclude that $D^R$ can be obtained as well within CSS approach when all terms are resummed up to their appropriate order.

The way the evolution is usually implemented within the CSS approach in the literature is described in Appendix~\ref{app:CSS}. Comparing Fig.~\ref{fig:kernelDR}b and Fig.~\ref{fig:css}a one sees a difference in the two approaches. 
This difference is apparent also in a numerical comparison of  Eq.~(\ref{eq:CSSD1}) with respect to Eq.~(\ref{eq:CSSD2}), as shown in Fig.~(\ref{fig:DLLorders}). 
The crucial point is that going from Eq.~(\ref{eq:CSSD1}) to Eq.~(\ref{eq:CSSD2}) requires that no higher order contributions from the  running of $\alpha_s$ are included and that the number of flavors included in the running of $\a_s(Q)$ and $\a_s(\mu_b)$ is the same.
In fact, even at one loop and taking $\as(M_Z)$ as a reference value for the running of the strong coupling, one has
($n_f[Q]$ is the number of active flavors at the scale $Q$)
\begin{align}
-\frac{\Gamma_0}{2\b_0(n_f[Q_i])}\ln
\le.\frac{\as(Q_i)}{\as(\m_b)}\ri|_{\rm 1-loop} &=
-\frac{\G_0}{2\b_0(n_f[Q_i])}\ln
\frac{1-\frac{\as(M_Z)}{4\pi}\b_0(n_f[\mu_b])\ln\frac{M_Z^2}{\m_b^2}}{1-\frac{\as(M_Z)}
{4\pi}\b_0(n_f[Q_i])\ln\frac{M_Z^2}{Q_i^2}}\ ,
\end{align}
and
\begin{align}
\le.-\frac{\G_0}{2\b_0(n_f[Q_i])}\ln(1-X) \ri|_{\rm 1-loop}= -\frac{\G_0}{2\b_0(n_f[Q_i])}\ln
\frac{1-\frac{\as(M_Z)}{4\pi}\b_0(n_f[Q_i])\ln\frac{M_Z^2}{\m_b^2}}{1-\frac{\as(M_Z)}
{4\pi}\b_0(n_f[Q_i])\ln\frac{M_Z^2}{Q_i^2}}
\,.
\end{align}
The difference can be appreciated from Figs.~\ref{fig:DLLorders}a and~\ref{fig:DLLorders}b. 
In order to clarify this problem we plot in Fig.~(\ref{fig:DLLorders}) the $D$ as in Eq.~(\ref{eq:CSSD1}) and also as in Eq.~(\ref{eq:CSSD2}), with several orders for the running of $\alpha_s$, starting from the usual value of $\a_s(M_Z=91.187~{\rm GeV)}=0.117$. 
It is straightforward to check that the solution provided by the $D^R$ is  stable, while the direct use of Eq.~(\ref{eq:CSSD1}) leads to undesired  divergent behavior for relatively low values of the impact parameter.

In our calculation we have implemented all decoupling corrections for $\as$ as given in~\cite{Grozin:2012ec,Larin:1994va,Chetyrkin:1997un,Chetyrkin:2000yt,Chetyrkin:2005ia,Schroder:2005hy} and we have set the mass thresholds at $m_c=1.2~{\rm GeV}$ and $m_b=4.2~{\rm GeV}$.
In other words, the implementation of $D^R$ takes into account the running of the coupling constant at the correct perturbative order and the decoupling of thresholds automatically.
The explicit formulas equivalent to Eqs.~(\ref{eq:CSSD1}) and~(\ref{eq:CSSD2}) at NLL and NNLL are given respectively in Eqs.~(\ref{eq:C2}) and~(\ref{eq:resummedD}).

We conclude from this analysis that the use of $D^R$ is by construction consistent with the considered perturbative order within the resummation scheme.
As a result, a direct implementation of Eq.~(\ref{eq:CSSD1}) with a running coupling at higher orders introduces higher order terms which spoil the convergence of the resummation for small values of $b$.
The same problem appears if instead of Eq.~(\ref{eq:CSSD1}) one considers its equivalent at NLL and NNLL, Eq.(\ref{eq:C2}). 
Within the standard CSS approach this issue is hidden behind the implementation of non-perturbative models, since the $b^*$ prescription washes it out. 
The consistent perturbative expansion performed with our $D^R$ allows us to separate more clearly the perturbative and non-perturbative regions of the evolution kernel.
The same conclusion can be established if one compares the direct use of Eq.~(\ref{eq:devolved0}) with a full running of the coupling constant with the $D^R$ given in Eq.~(\ref{eq:resummedD}), which is consistent with the logarithmic accuracy within the resummation scheme.

\section{Comparison with CSS Approach}
\label{sec:comparison}

\begin{figure}[t]
\begin{center}
\includegraphics[width=0.45\textwidth]{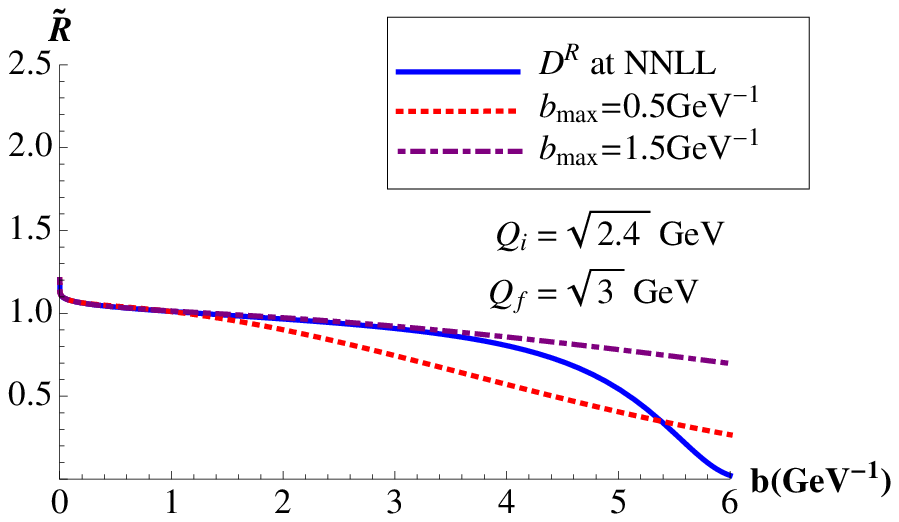}
\quad\quad\quad
\includegraphics[width=0.45\textwidth]{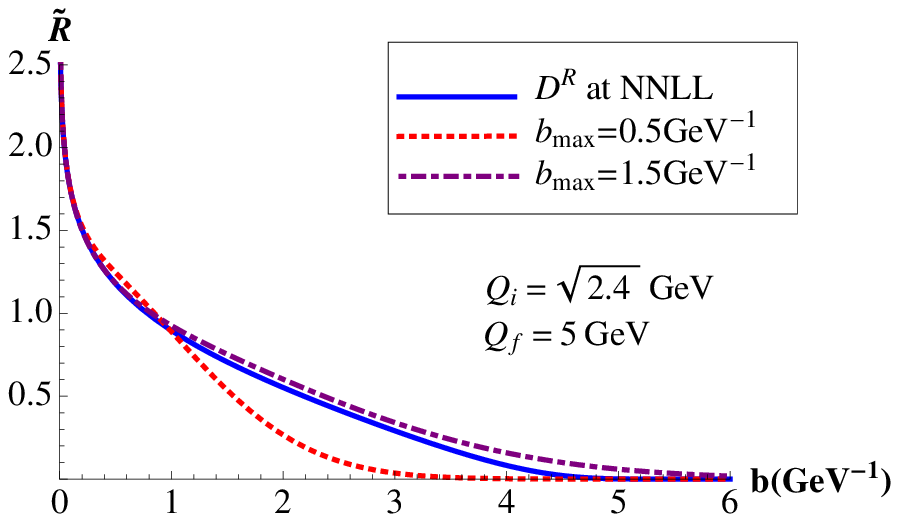}
\\
\hspace{0cm}(a)\hspace{9cm}(b)\hspace{6cm}
\\
\includegraphics[width=0.45\textwidth]{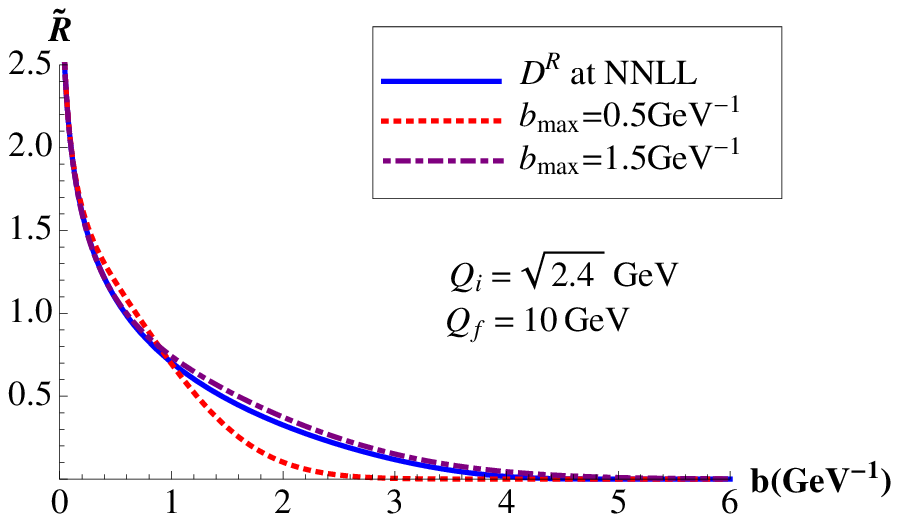}
\quad\quad\quad
\includegraphics[width=0.45\textwidth]{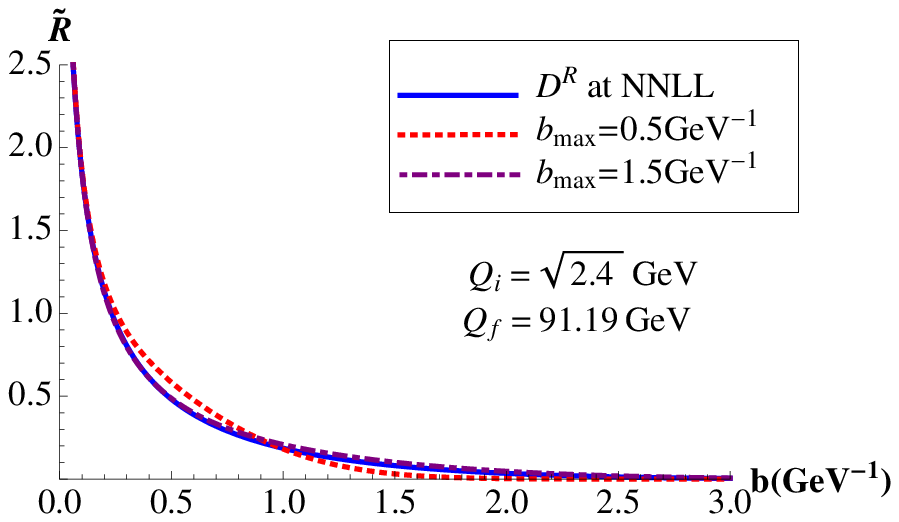}
\\
\hspace{0cm}(c)\hspace{9cm}(d)\hspace{6cm}
\end{center}
\caption{\it
Evolution kernel from $Q_i=\sqrt{2.4}~{\rm GeV}$ up to $Q_f=\{\sqrt{3}\,,5\,,10\,,91.19\}~{\rm GeV}$ using ours and CSS approaches, both at NNLL.
}
\label{fig:kernelDRandCSS}
\end{figure}

In this section we consider our resummed expression for the evolution kernel given in Eq.~(\ref{eq:Rc}) and compare it with the one within CSS approach (which for completeness is outlined in Appendix~\ref{app:CSS}) given in Eq.~(\ref{eq:kernelcss}).
Notice that the main difference between both lies in the separation between the perturbative and non-perturbative regions.
While in our case we clearly separate both regimes, achieving a completely perturbative expression for the kernel in the small $b$ region, however within the CSS approach the two contributions overlap in the perturbative domain.
In other words, the $b^*$ prescription implements a smooth cutoff between the perturbative and non-perturbative domains.
Using the results of the previous section, within our approach we have
\begin{align}
D(b;Q_i) &= D^R(b;Q_i)\theta(b_c-b) + D^{NP}(b;Q_i)\theta(b-b_c)
\,,
\end{align}
while for CSS (with BLNY model) we have
\begin{align}
D(b;Q_i) &= 
D(b^*;\m_b) + \int_{\m_{b^*}}^{Q_i} \frac{d\bar\m}{\bar\m} \G_{\rm cusp} 
+ \frac{1}{4} g_2 b^2
\nn\\
&= 
D^R(b^*;Q_i) + \frac{1}{4} g_2 b^2
\,.
\end{align}
Definitely within the CSS approach the non-perturbative model has some effect also in the small $b$ region, and at the same time, the perturbative contribution is not completely parameter free since it is cut off by the implementation of the $b^*$ prescription and depends on the value of that parameter.

\begin{figure}[t]
\begin{center}
\includegraphics[width=0.44\textwidth]{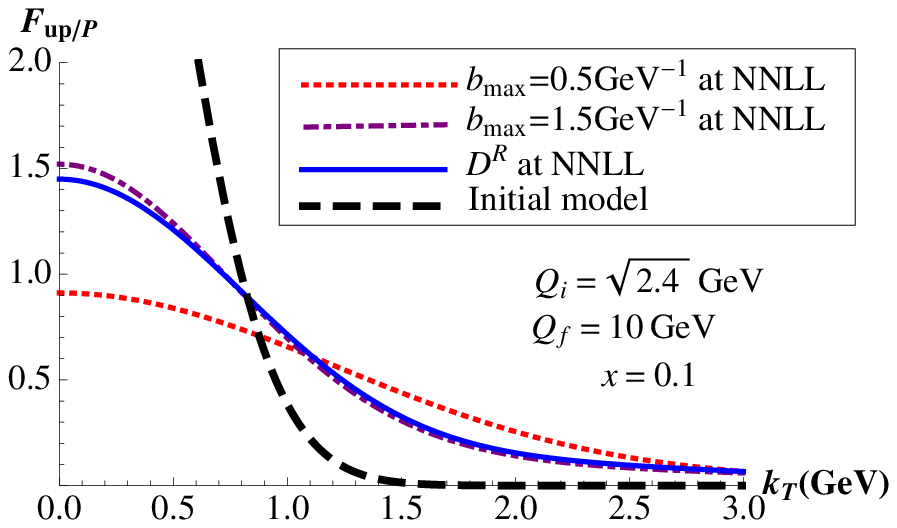}
\\
\hspace{0cm}(a)\hspace{9cm}
\\
\includegraphics[width=0.45\textwidth]{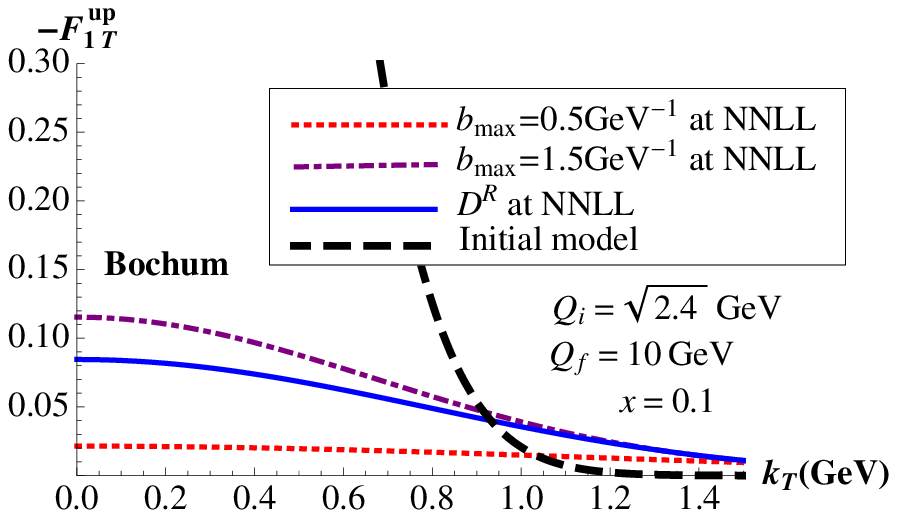}
\quad\quad\quad
\includegraphics[width=0.45\textwidth]{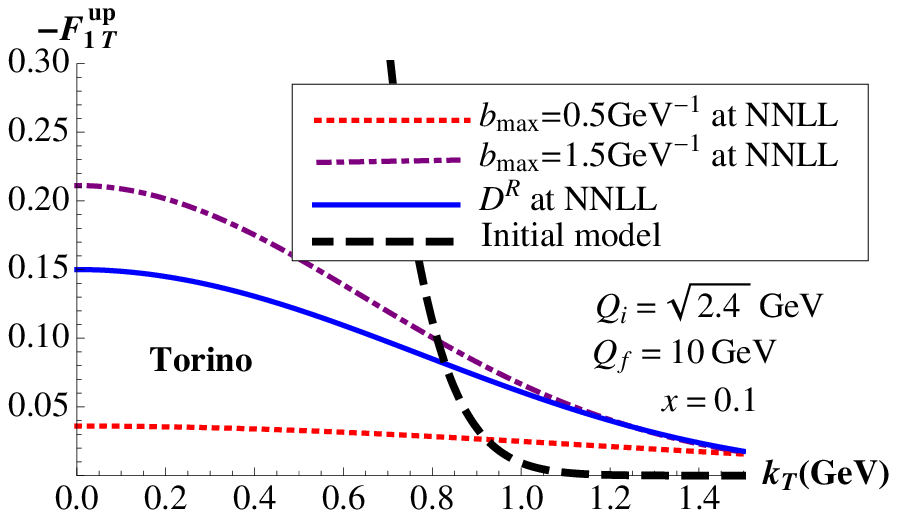}
\\
\hspace{0cm}(b)\hspace{9cm}(c)\hspace{6cm}
\end{center}
\caption{\it
Up quark unpolarized TMDPDF and Sivers function (Bochum and Torino fits) evolved from $Q_i=\sqrt{2.4}~{\rm GeV}$ up to $Q_f=10~{\rm GeV}$ with different approaches to the evolution kernel.
Black line stands for the input Gaussian model and the rest for the evolved TMD either with CSS or our approaches.}
\label{fig:evolvedtmds}
\end{figure}

In order to perform the resummation of large logarithms consistently up to $\textrm{N}^i\textrm{LL}$ order (or $\textrm{N}^{i-1}\textrm{LO}$ in RG-improved perturbation theory) one needs the ingredients shown in Tables~\ref{tab:res}--\ref{tab:rescss},
where $n$ stands for the order of the perturbative QCD calculation, $k$ for the power of the large logarithms encountered in those calculations and $i$ for the accuracy of the resummed logarithms.
In our approach one takes the resummed series in Eq.~(\ref{eq:resummedD}) up to the corresponding order $i$.
In~\cite{Aybat:2011zv,Aybat:2011ge,Anselmino:2012aa} the cusp anomalous dimension $\G_{\rm cusp}$ was not implemented up to 2-loop order, as it should be to get a complete NLL result.
In Figs.~\ref{fig:kernelDRandCSS} and~\ref{fig:evolvedtmds} we have implemented $\g_F$, $\G_{\rm cusp}$ and $D$ consistently within the CSS approach to achieve the $\textrm{N}^i\textrm{LL}$ accuracy.

\begin{table}[h]
\begin{center}
\begin{tabular}{c|c|ccc}
Order			&	Accuracy $\sim \as^n L^k$			&	$\g^V$	&	$\G_{\rm cusp}$		&	$D^R$
\\
\hline
$\textrm{N}^i\textrm{LL}$		&	$n+1-i\leq k \leq 2n \,\,(\as^{i-1})$ &	$\as^{i}$	&	$\as^{i+1}$			&	$(\as/(1-X))^i$
\end{tabular}
\end{center}
\caption{
Approximation schemes for the evolution of the TMDs with $D^R$, where $L=\ln(Q_f^2/Q_i^2)$ and $\as^i$ indicates the order of the perturbative expansion.
}
\label{tab:res}
\end{table}

In Fig.~\ref{fig:kernelDRandCSS} we compare our approach to the evolution kernel with CSS, both at NNLL.
On one hand, as already mentioned, it is clear that our approach can be applied only when the contribution of the non-perturbative large $b$ region is negligible, as it is the case for large enough $Q_f$.
On the other hand, since our expression for the evolution kernel is parameter free up to $b_c\sim4~{\rm GeV}^{-1}$, from all plots one can deduce that $b_{\rm max}=1.5~{\rm GeV}^{-1}$ gives better results in that region.
In fact, this is the value that was found in~\cite{Konychev:2005iy} by fitting experimental data.
Previous fits did not consider $b_{\rm max}$ as a free parameter, but rather set it to $0.5~{\rm GeV}^{-1}$ right from the start, fitting just the rest of the parameters of the non-perturbative model.

\begin{figure}[t]
\begin{center}
\includegraphics[width=0.45\textwidth]{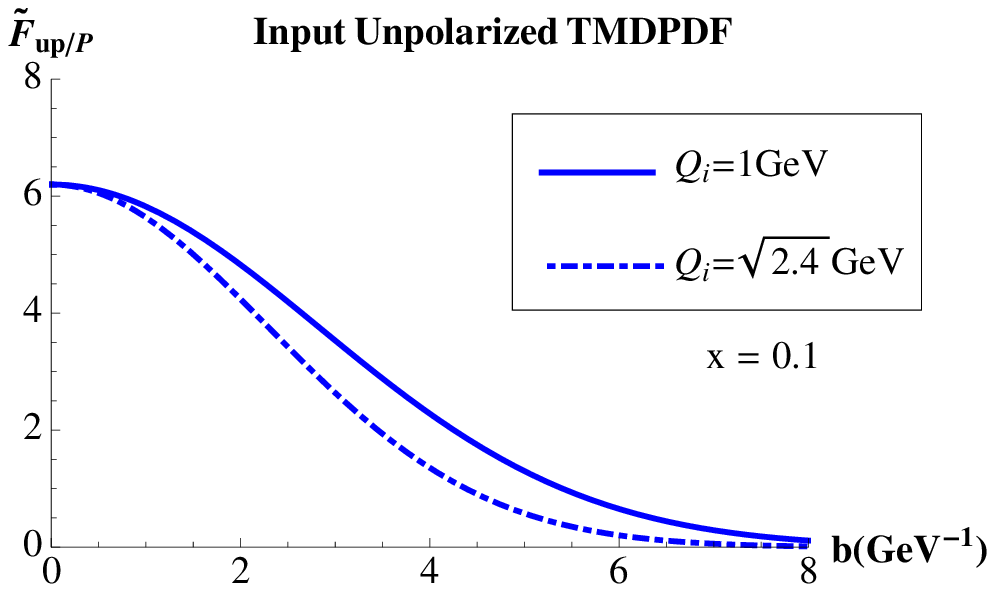}
\quad\quad\quad
\includegraphics[width=0.45\textwidth]{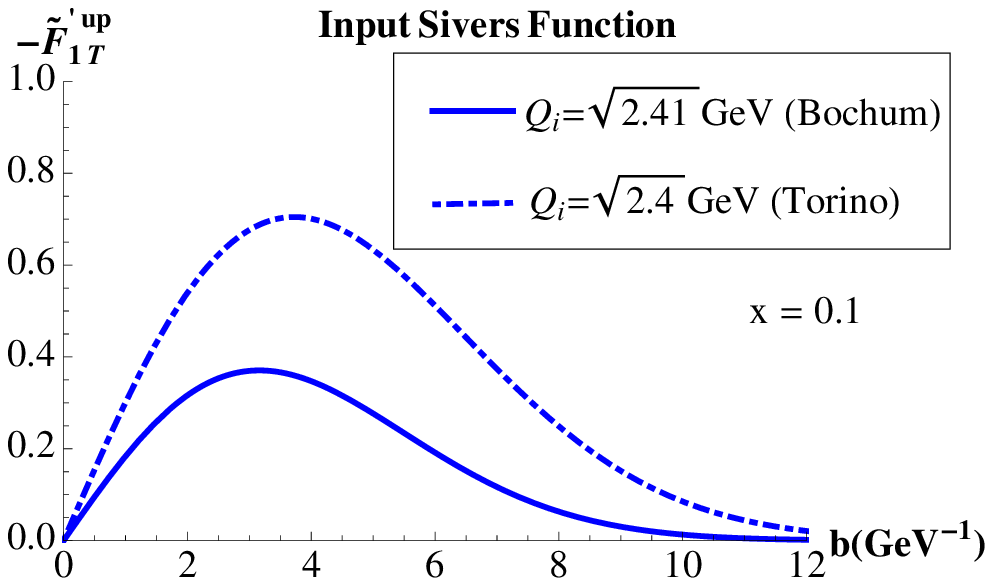}
\\
\hspace{0cm}(a)\hspace{9cm}(b)\hspace{3cm}
\end{center}
\caption{\it
a) Input unpolarized up-quark TMDPDF for $Q_i=\{1,\sqrt{2.4}\}~{\rm GeV}$~\cite{Anselmino:2005nn,Schweitzer:2010tt}.
b) Input Sivers function following Bochum~\cite{Collins:2005ie} and Torino~\cite{Anselmino:2008sga} fits.
}
\label{fig:inputtmd}
\end{figure}

To illustrate the application of the evolution kernel within our formalism  and CSS one, we consider an input functions for the unpolarized TMDPDF~\cite{Anselmino:2005nn,Schweitzer:2010tt} and the Sivers function~\cite{Collins:2005ie,Anselmino:2008sga}, shown in Fig.~\ref{fig:inputtmd}.
The unpolarized quark-TMDPDF at low energy is modeled as a Gaussian,
\begin{align}
\tilde{F}_{up/P}(x,b;Q_i) &=
f_{up/P}(x;Q_i)\exp[-\s b_T^2]
\,,
\end{align}
with $\s=0.25/4\,{\rm GeV}^{2}$ for $Q_i=1~{\rm GeV}$~\cite{Anselmino:2005nn} and $\s=0.38/4\,{\rm GeV}^{2}$ for $Q_i=\sqrt{2.4}~{\rm GeV}$~\cite{Schweitzer:2010tt}, and $f_{up/P}$ the up-quark integrated PDF, which has been taken from the MSTW data set~\cite{Martin:2009iq}.
The Sivers function at low energy is modeled following what are called the ``Bochum''~\cite{Collins:2005ie} and ``Torino''~\cite{Anselmino:2008sga} fits in~\cite{Aybat:2011ge}.
The evolved TMDs using our and CSS approaches at NNLL are shown in Fig.~\ref{fig:evolvedtmds}.
The slight difference between our kernel and the one of CSS with $b_{\rm max}=1.5~{\rm GeV}^{-1}$ in Fig.~\ref{fig:kernelDRandCSS}c is washed out in the case of the unpolarized TMDPDF, since the input function is narrower.
For the Sivers function, which is wider at the initial scale, we see a bigger difference.
In any case, given the fact that in this kinematical setup the resummed expression for the evolution kernel in Eq.~(\ref{eq:Rc}) is parameter free and convergent, as it is clear from Fig.\ref{fig:kernelDR}c, solid blue curves should be considered as the most accurate ones.

\section{Conclusions}
\label{sec:conclusions}

We have argued that the evolution of leading twist TMDs, both for polarized and unpolarized ones, is driven by the same kernel, which can be obtained up to NNLL accuracy by using the currently known results for the cusp anomalous dimension and the QCD $\beta$-function. 
For completeness, we have provided as well an expression for the evolution kernel at NNNLL.

The evolution kernel, as a function of the impact parameter $b$, can be obtained in the perturbative region without introducing any model dependence, and the resummation can be performed up to any desired logarithmic accuracy.
This resummation can be done either by solving a recursive differential equation or by properly implementing the running of the strong coupling with renormalization scale within the standard CSS approach. 
In both cases we obtained identical results. 
This fact is actually not surprising. 
The definition of (unpolarized) TMDPDF, both in the EIS~\cite{Echevarria:2012js} and JCC~\cite{Collins:2011zzd} approaches has been shown to be equivalent~\cite{Echevarria:2012js,Collins:2012uy}. 
If the resummation of the large logarithms is performed properly and consistently (in terms of logarithmic accuracy) then the final results for the evolved TMDs should agree as well. 
We consider this agreement as one of the major contributions of this work as it unifies a seemingly different approaches to TMD theory and phenomenology.

As already mentioned, one of the contributions of  this paper is to give a parameter free expression for the evolution kernel by using the highest order perturbatively calculable known ingredients, which  is valid only within the perturbative domain in the impact parameter space.
On the other side, within the CSS approach there is an overlap between the perturbative and non-perturbative contributions to the evolution kernel, due to the implementation of a smooth cutoff through the $b^*$ prescription.
Comparing both approaches, we conclude that $b_{\rm max}=1.5~{\rm GeV}^{-1}$, which is more consistent with our results (see Fig.~\ref{fig:kernelDRandCSS}), gives a better description of the perturbative region within the CSS approach, as was actually found by phenomenological fits in~\cite{Konychev:2005iy}.

We have studied under which kinematical conditions the non-perturbative contribution to the kernel is negligible, and hence our approximate expression for the kernel (in Eq.~(\ref{eq:Rc})) can be applied without recurring to any model for all practical purposes.
Given an initial scale $Q_i=\sqrt{2.4}~{\rm GeV}$, at which one would like to extract the low energy models for TMDs, if the final scale is $Q_f\geq 5~{\rm GeV}$, then the effects of non-perturbative physics are washed out, as is shown in Fig.~\ref{fig:kernelDR}. 
In this case, all the model dependence is restricted to the low energy functional form of TMDs to be extracted by fitting to data.

Thus, phenomenologically, the major point of our work is to provide a scheme optimized for the extraction of TMDs from data. 
Assuming that low energy models are to be extracted at scale $Q_i \sim 1-2~{\rm GeV}$, if data are selected with $Q_f > 5~{\rm GeV}$, then the evolution is in practice parameter free.
For instance, COMPASS, Belle or BaBar facilities can perfectly fulfill these requirements.

Finally, the definition of quark-TMDPDFs given in Eq.~(\ref{eq:tmddef}) and the new approach to determine the evolution kernel can be extended to gluon-TMDPDFs~\cite{Mulders:2000sh} and quark/gluon TMDFFs.
This approach can be applied as well to the evolution kernel of the complete hadronic tensor $\tilde M$ (built with two TMDs).
As a future application one might consider the use of low energy TMDs as input hadronic matrix elements for large-energy hadronic colliders, where the evolution kernels could be implemented in a model independent way, leaving all the non-perturbative contributions to the TMDs themselves.

\section*{Acknowledgements}

This work is supported by the Spanish MEC, FPA2011-27853-CO2-02.
We would like to thank Umberto D'Alesio, V. Braun, M. Diehl, A. Metz and J. Zhou for useful discussions.
M.G.E. is supported by the PhD funding program of the Basque Country Government.
A.I. and A.S. are supported by BMBF (06RY9191).
I.S. was partly supported by the Ram\'on y Cajal Program.

\appendix

\section{CSS Approach to the Evolution of TMDs}
\label{app:CSS}

In various works following Collins' approach to TMDs~\cite{Anselmino:2012aa,Collins:2011zzd,Aybat:2011ge}, large $L_\perp$ logarithms in the $D$ term of the evolution kernel in Eq.~(\ref{eq:tmdkernel}) were resummed using the CSS approach~\cite{Collins:1984kg}, which, for the sake of completeness and comparison, we explain below.

\begin{figure}[h]
\begin{center}
\includegraphics[width=0.32\textwidth]{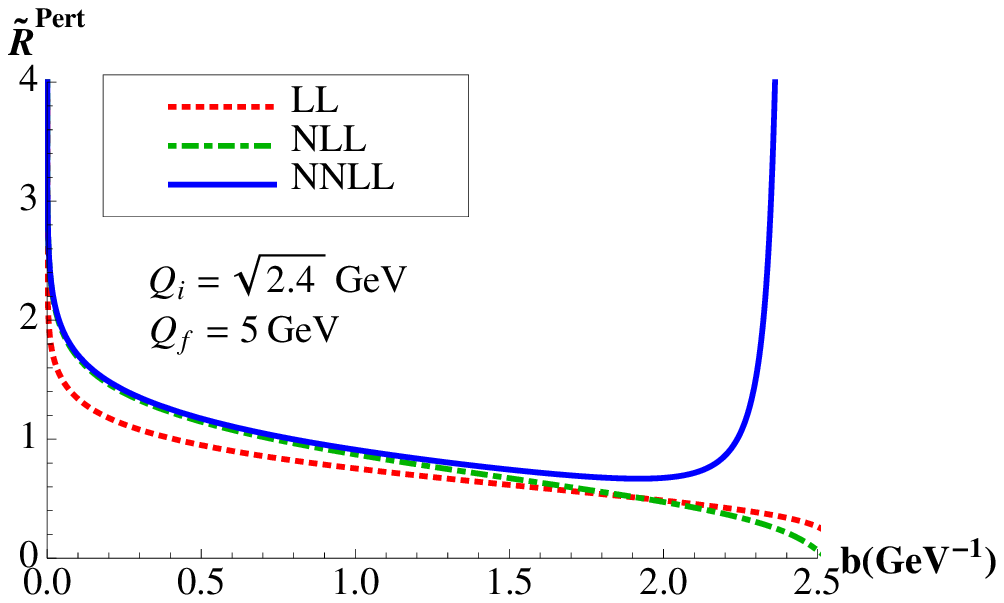}
\includegraphics[width=0.32\textwidth]{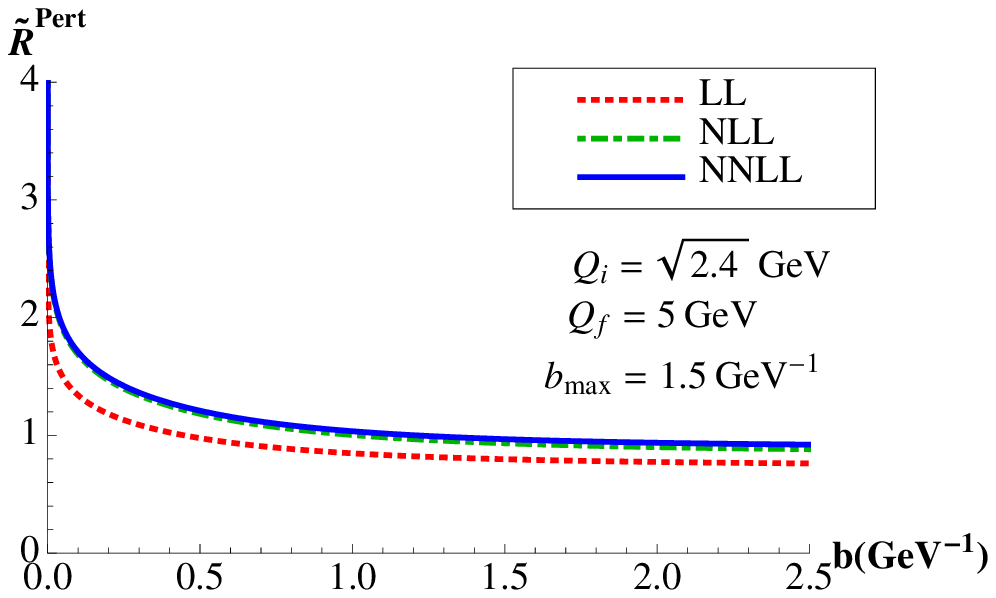}
\includegraphics[width=0.32\textwidth]{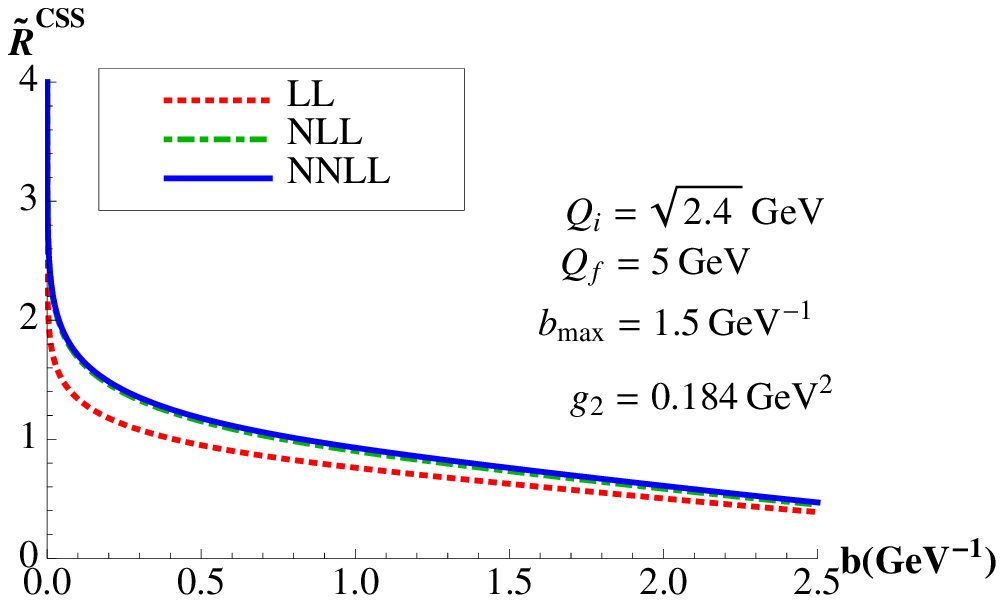}
\\
\hspace{0cm}(a)\hspace{6cm}(b)\hspace{5.5cm}(c)\hspace{1cm}
\end{center}
\caption{\it
Evolution kernel from $Q_i=\sqrt{2.4}~{\rm GeV}$ up to $Q_f=5~{\rm GeV}$  using RG-evolution in~Eq.~(\ref{eq:devolved}) to resum the $D$ term.
The resummation accuracy is given in table~\ref{tab:rescss}.
(a) With $\m_b = 2e^{-\g_E}/b$ the Landau pole appears clearly. 
(b) With $\m_{b^*} = 2e^{-\g_E}/{b^*}$ and $b_{\rm max}=1.5~{\rm GeV}^{-1}$ to avoid hitting the Landau pole.
(c) Adding the BLNY non-perturbative model to recover the information at large $b$.
}
\label{fig:css}
\end{figure}

First, the $D$ term is resummed using its RG-evolution in Eq.~(\ref{eq:drg}),
\begin{align}\label{eq:devolved}
D\le(b;Q_i\ri) &=
D\le(b;\m_b\ri)
+ \int_{\m_b}^{Q_i}\frac{d\bar\m}{\bar\m} \G_{\rm cusp}
\,,
\end{align}
where large $L_\perp$ logarithms in the $D$ term on the right hand side are cancelled by choosing $\m_b = 2e^{-\g_E}/b$.
Thus, they are resummed by evolving  $D$  from $\m_b$ to $Q_i$.
However, since we need to Fourier transform back to momentum space, at some value of $b$ the effective coupling $\as(\m_b)$ will hit the Landau pole.
In Fig.~\ref{fig:css}(a) we can see the evolution kernel $\tilde{R}(b;Q_i=\sqrt{2.4}~{\rm {GeV}},Q_f=5~{\rm{GeV}})$ where we have used~Eq.~(\ref{eq:devolved}) to resum the $L_\perp$ logarithms in $D$ and the appearance of the Landau pole is manifest.
The breakdown of the perturbative series is driven by the running coupling $\as(\m_b)$, when $\m_b$ is sufficiently small.

\begin{table}[h]
\begin{center}
\begin{tabular}{c|c|ccc}
Order			&	Accuracy $\sim \as^n L^k$			&	$\g^V$	&	$\G_{\rm cusp}$		&	$D$
\\
\hline
$\textrm{N}^i\textrm{LL}$		&	$n+1-i\leq k \leq 2n \,\,(\as^{i-1})$ &	$\as^{i}$	&	$\as^{i+1}$			&	$\as^i$
\end{tabular}
\end{center}
\caption{Approximation schemes for the evolution of TMDs with CSS approach, where $L=\ln(Q_f^2/Q_i^2)$ and $\as^i$ indicates the order of the perturbative expansion.}
\label{tab:rescss}
\end{table}

In order to avoid this issue, CSS did not actually introduce a sharp cut-off but a smoothed one defined as $b^*=b/\sqrt{1+(b/b_{\rm max})^2}$.
Obviously $b^*$ cannot exceed $b_{\rm max}$ and the effective coupling $\as(\m_{b^*})$ does not hit the Landau pole.
As is shown in Fig.~\ref{fig:css}(b), the kernel saturates and does not present any uncontrolled behavior.
It is also worth noticing that comparing Fig.~\ref{fig:css}(a) with Fig.~\ref{fig:css}(b), we see that the implementation of the $b^*$ prescription has some appreciable effect in the perturbative region, which now depends on this parameter.

The lost information due to the cutoff is recovered by adding a non-perturbative model that has to be extracted from experimental data of a measured cross section.
This model not only gives the proper information in the non-perturbative region, but also restores the correct shape of the kernel within the perturbative domain, which was affected by the $b^*$ prescription.
When implementing, for example, the Brock-Landry-Nadolsky-Yuan (BLNY )model the evolution kernel can be written as
\begin{align}\label{eq:kernelcss}
\tilde{R}^{\rm CSS}(b;Q_i,Q_f) &=
\exp\le\{
\int_{Q_i}^{Q_f} \frac{d\bar\m}{\bar\m} \g_F\le(\as(\bar\m),\ln\frac{Q_f^2}{\bar\m^2} \ri)
\ri\}
\le(\frac{Q_f^2}{Q_i^2}\ri)^
{-\le[D\le(b^*;Q_i\ri)+\frac{1}{4}g_2 b^2\ri]}
 \,,
\end{align}
where $D(b^*;Q_i)$ is resummed using Eq.~(\ref{eq:devolved}).
In this model $g_2=0.68~{\rm GeV}^2$ for $b_{\rm max}=0.5~{\rm GeV}^{-1}$~\cite{Landry:2002ix} and $g_2=0.184~{\rm GeV}^2$ for $b_{\rm max}=1.5~{\rm GeV}^{-1}$~\cite{Konychev:2005iy}.
From the theoretical point of view these two choices are legitimate and they can be used to define the model dependence of the final result. 
However considering $b_{\rm max}$ as a fitting parameter the choice of
$b_{\rm max}=1.5~{\rm GeV}^{-1}$ should be preferred according to Ref.~\cite{Konychev:2005iy}.
Fig.~\ref{fig:css}(c) shows the complete evolution kernel with the CSS approach, Eq.~(\ref{eq:kernelcss}), while implementing the BLNY model.

\section{Derivation of $D^R$ up to NNNLL}
\label{sec:uyy}

Below we provide the details of the derivation of $D^R$ within the CSS formalism, i.e., solving Eq.~(\ref{eq:devolved0}). 
Using the perturbative expansion of $\G_{\rm cusp}(\as)$ and $\b(\as)$ one can write,
\begin{align}\label{eq:C2}
&\int_{\m_b}^{Q_i}  d(\ln\m) \G_{\rm cusp} =
\int_{\as(\m_b)}^{\as(Q_i)}  d\a\; \frac{\G_{\rm cusp}(\a)}{\b(\a)}
\nn\\
&=\int_{\as(\m_b)}^{\as(Q_i)}  d\a\;
\le\{
\frac{-\Gamma_0}{2 \alpha \beta_0}+
\frac{\Gamma_0\beta_1-\Gamma_1 \beta_0}{8 \pi  \beta_0^2}+
\frac{\alpha  \left(-\beta_0^2 \Gamma_2+\beta_0 \beta_1 \Gamma_1+\beta_0 \beta_2 \Gamma_0-\beta_1^2 \Gamma_0\right)}{32 \pi ^2 \beta_0^3}
\ri.
\nn\\
&\le.
+
\frac{\alpha ^2 \left( -\beta_0^3 \Gamma_3+\beta_0^2
 \beta_1 \Gamma_2+\beta_0^2 \beta_2\Gamma_1+\beta_0^2 \beta_3\Gamma_0-\beta_0
   \beta_1^2 \Gamma_1-2 \beta_0 \beta_1\beta_2 \Gamma_0+\beta_1^3 \Gamma_0\right)}{128 \pi ^3\beta_0^4}
\ri\}
\nn\\
&=
\frac{-\G_0}{2  \b_0}\ln\frac{\as(Q_i)}{\as(\m_b)}+
\big[\as(Q_i)-\as(\m_b)\big]\frac{\G_0\b_1-\G_1 \b_0}{8 \pi  \b_0^2}+
\big[\as^2(Q_i)-\as^2(\m_b)\big]\frac{ \left(-\beta_0^2 \Gamma_2+\beta_0 \beta_1 \Gamma_1+\beta_0 \beta_2 \Gamma_0-\beta_1^2 \Gamma_0\right)}{64 \pi ^2 \beta_0^3}\nn \\
&+
\big[\as^3(Q_i)-\as^3(\m_b)\big]\frac{\left( -\beta_0^3 \Gamma_3+\beta_0^2
 \beta_1 \Gamma_2+\beta_0^2 \beta_2\Gamma_1+\beta_0^2 \beta_3\Gamma_0-\beta_0
   \beta_1^2 \Gamma_1-2 \beta_0 \beta_1\beta_2 \Gamma_0+\beta_1^3 \Gamma_0\right)}{384 \pi ^3\beta_0^4}
\,.
\end{align}
Then in the equation above we use the running of the strong coupling to expand $\as(\m_b)$ in terms of $\as(Q_i)$, 
\begin{align}
\a_s(\mu_b)&=
\a_s(Q_i)\frac{1}{1-X} - \a_s^2(Q_i)\frac{\b_1 \ln(1-X)}{4\pi (1-X)^2 \b_0}-
\a_s^3(Q_i)\frac{\left(-X \beta_0 \beta_2+
  \beta_1^2 (X-\ln^2(1-X)+ \ln (1-X))\right)}{16 \pi ^2 (1-X)^3
   \beta_0^2}
\nn\\%
&
-\a_s^4(Q_i)\frac{\left(\beta_1^3 \left(X^2+2 \ln ^3(1-X)-5 \ln
   ^2(1-X)-4 X \ln (1-X)\right)\right)}{128 \pi ^3 (X-1)^4
   \beta_0^3}
\nn\\
&-\a_s^4(Q_i)\frac{\left(+2\beta_0\beta_1\beta_2 ((2 X+1) \ln (1-X)-(X-1) X)+(X-2) X
   \beta_0^2 \beta_3\right)}{128 \pi ^3 (X-1)^4\beta_0^3}
\label{eq:C3}
\end{align}
and implement it up to the appropriate order in $\a_s(Q_i)$.
In order to finally get $D^R$ at NNLL one should consider also the term $D(b;\m_b)$ in Eq.~(\ref{eq:devolved0}), which at second order does not vanish due to the presence of the finite $d_2(0)$ term,
\begin{align}
D^{(2)}(b;\mu_b)= 
d_2(0) \le(\frac{\a_s(\m_b)}{4\pi}\ri)^2=d_2(0) \frac{a^2}{(1-X)^2}
\,,
\end{align}
with $a=\as(Q_i)/4\pi$.
Inserting this result in Eq.~(\ref{eq:C2}) and the expansion in Eq.~(\ref{eq:C3}) up to order $\a_s^2(Q_i)$, one gets our Eq.~(\ref{eq:resummedD}).

Finally, for completeness and future reference, we provide also $D^R$ at NNNLL,
\begin{align}
\label{eq:C4}
D^{R(3)}&= \frac{a^3}{(1-X)^3}\left(d_3(0)-2d_2(0)\frac{\b_1}{\b_0}\ln(1-X)+D^{R(3)}_\Gamma\right)\ ,
\nn \\
D^{R(3)}_\Gamma&=-\frac{1}{12\beta_0^4}\left[\beta_0^2 \left(2 \Gamma_2 \beta_1
   \left(X \left(X^2-3 X+3\right)+3 \ln (1-X)\right)+X^2 (2 \Gamma_1 (X-3)
  \beta_2+\Gamma_0 (2 X-3)\beta_3   )\right)\right.
\nn \\
&
-2 \beta_0\beta_1
   \left(\Gamma_1\beta_1 \left((X-3) X^2+3 \ln^2(1-X)\right)+\Gamma_0 X \beta_2 (X (2 X-3)-3 \ln
   (1-X))\right)-2 \Gamma_3 X \left(X^2-3 X+3\right) \beta_0^3
\nn  \\ & 
\left.
+\Gamma_0 \beta_1^3 \left(X^2 (2 X-3)+2
   \ln^3(1-X)-3 \ln^2(1-X)-6 X \ln (1-X)\right)\right]\ .
\end{align}

\section{Evolution of the Hard Matching Coefficient}
\label{sec:app}

The evolution of the hard matching coefficient $C_V$, where $H=|C_V|^2$, is given by
\begin{align}\label{gammaV}
\frac{d}{d\ln\mu}\ln C_V(Q^2/\m^2) &=
\g_{C_V}\le(\as(\m),\ln\frac{Q^2}{\m^2} \ri)
\,,
\quad\quad
\g_{C_V} =
\Gamma_{\rm cusp}(\alpha_s)\,\ln\frac{Q^2}{\mu^2}
+ \g^V(\as)
\,,
\end{align}
where the cusp term is related to the evolution of the Sudakov double logarithms and the remaining term with the evolution of single logarithms.
The exact solution of this equation is
\begin{align}\label{CVsol}
C_V(Q^2/\m_f^2) &=
C_V(Q^2/\m_i^2)\,
\exp\le[ \int_{\m_i}^{\m_f} \frac{d\bar\m}{\bar\m}\,
\g_{C_V}\le(\as(\bar\m),\ln\frac{Q^2}{\bar\m^2} \ri)\ri]
\nn\\
&=
C_V(Q^2/\m_i^2)\,
\exp\le[ \int_{\as(\m_i)}^{\as(\m_f)} \frac{d\bar\as}{\b(\bar\as)}\,
\g_{C_V}\le(\bar\as\ri)\ri]
\,,
\end{align}
where we have used that $d/d\ln\m=\b(\as)\,d/d\as$, where
$\b(\as)=d\as/d\ln\m$ is the QCD $\b$-function.

Below we give the expressions for the anomalous dimensions and the QCD $\beta$-function, in the $\overline{{\rm MS}}$ renormalization scheme.
We use the following expansions:
\begin{align}
\G_{\rm cusp} &=
\sum_{n=1}^{\infty} \G_{n-1} \le( \frac{\as}{4\pi}\ri)^n
\,,
\quad
\g^V = \sum_{n=1}^{\infty} \g^V_{n} \le( \frac{\as}{4\pi}\ri)^n
\,,
\quad
\b= -2\as \sum_{n=1}^\infty \b_{n-1} \left( \frac{\as}{4\pi} \right)^n
\,.
\end{align}
The coefficients for the cusp anomalous dimension $\G_{\rm cusp}$ are
\begin{align}
\G_0 &=
4 C_F
\,,
\nn\\
\G_1 &=
4 C_F \le[ \le( \frac{67}{9} - \frac{\pi^2}{3} \ri) C_A - \frac{20}{9}\,T_F n_f \ri]
\,,
\nn\\
\G_2 &=
4 C_F \le[ C_A^2 \left( \frac{245}{6} - \frac{134\pi^2}{27}
+ \frac{11\pi^4}{45} + \frac{22}{3}\,\zeta_3 \right)
+ C_A T_F n_f  \left( - \frac{418}{27} + \frac{40\pi^2}{27}
- \frac{56}{3}\,\zeta_3 \right)
\ri.
\nn\\
&\le.
+ C_F T_F n_f \left( - \frac{55}{3} + 16\zeta_3 \right)
- \frac{16}{27}\,T_F^2 n_f^2 \ri]
\,.
\end{align}
The anomalous dimension $\gamma^V$ can be determined up to three-loop order from the partial three-loop expression for the on-shell quark form factor in
QCD.
We have
\begin{align}
\g_0^V &=
-6 C_F
\,,
\nn\\
\g_1^V &=
C_F^2 \left( -3 + 4\pi^2 - 48\zeta_3 \right)
+ C_F C_A \left( - \frac{961}{27} - \frac{11\pi^2}{3} + 52\zeta_3 \right)
+ C_F T_F n_f \left( \frac{260}{27} + \frac{4\pi^2}{3} \right)
\,,
\nn\\
\gamma_2^V &=
C_F^3 \left( -29 - 6\pi^2 - \frac{16\pi^4}{5}
- 136\zeta_3 + \frac{32\pi^2}{3}\,\zeta_3 + 480\zeta_5 \right)
\nn\\
&
+ C_F^2 C_A \left( - \frac{151}{2} + \frac{410\pi^2}{9}
+ \frac{494\pi^4}{135} - \frac{1688}{3}\,\zeta_3
- \frac{16\pi^2}{3}\,\zeta_3 - 240\zeta_5 \right)
\nn\\
&
+ C_F C_A^2 \left( - \frac{139345}{1458} - \frac{7163\pi^2}{243}
- \frac{83\pi^4}{45} + \frac{7052}{9}\,\zeta_3
- \frac{88\pi^2}{9}\,\zeta_3 - 272\zeta_5 \right)
\nn\\
&
+ C_F^2 T_F n_f \left( \frac{5906}{27} - \frac{52\pi^2}{9}
- \frac{56\pi^4}{27} + \frac{1024}{9}\,\zeta_3 \right)
\nn\\
&
+ C_F C_A T_F n_f \left( - \frac{34636}{729}
+ \frac{5188\pi^2}{243} + \frac{44\pi^4}{45} - \frac{3856}{27}\,\zeta_3
\right)
+ C_F T_F^2 n_f^2 \left( \frac{19336}{729} - \frac{80\pi^2}{27}
- \frac{64}{27}\,\zeta_3 \right)
\,.
\end{align}
Finally, the coefficients for the QCD $\beta$-function are
\begin{align}
\b_0 &=
\frac{11}{3}\,C_A - \frac43\,T_F n_f
\,,
\nn\\
\b_1 &=
\frac{34}{3}\,C_A^2 - \frac{20}{3}\,C_A T_F n_f
- 4 C_F T_F n_f
\,,
\nn\\
\b_2 &=
\frac{2857}{54}\,C_A^3 + \left( 2 C_F^2
- \frac{205}{9}\,C_F C_A - \frac{1415}{27}\,C_A^2 \right) T_F n_f
+ \left( \frac{44}{9}\,C_F + \frac{158}{27}\,C_A \right) T_F^2 n_f^2
\,,
\nn\\
\b_3 &=
\frac{149753}{6} + 3564\zeta_3
- \left( \frac{1078361}{162} + \frac{6508}{27}\,\zeta_3 \right) n_f
+ \left( \frac{50065}{162} + \frac{6472}{81}\,\zeta_3 \right) n_f^2
+ \frac{1093}{729}\,n_f^3
\,,
\end{align}
where for $\b_3$ we have used $N_c=3$ and $T_F=\frac{1}{2}$.

\bibliography{references}

\end{document}